\documentclass{article}%
\usepackage{latexsym}
\usepackage{amssymb}
\usepackage{amsmath}
\usepackage{url}
\usepackage{cite}
\usepackage{amsfonts}
\usepackage{graphicx}%
\begin{document}

\title{ Fly-automata for checking \\MSO$_{2}$ graph properties.}
\author{Bruno Courcelle\\LABRI\thanks{This work has been supported by the French National Research
Agency (ANR) in the IdEx Bordeaux program "Investments for the future", CPU,
ANR-10-IDEX-03-02.}, Bordeaux University, Talence, France}
\maketitle

\begin{abstract}
A more descriptive but too long title would be : \emph{Constructing
fly-automata to check properties of graphs of bounded tree-width expressed by
monadic second-order formulas written with edge quantifications}.\ Such
properties are called MSO$_{2}$ in short. \emph{Fly-automata} (\emph{FA}) run
bottom-up on terms denoting graphs and\emph{\ }compute "on the fly" the
necessary states and transitions instead of\ looking into huge,\ actually
unimplementable tables.\ In previous works, we have constructed FA that
process terms denoting graphs of bounded clique-width, in order to check their
monadic second-order (MSO) properties (expressed by formulas without edge
quantifications).\ Here, we adapt these FA to \emph{incidence graphs}, so that
they can check MSO$_{2}$ properties of graphs of bounded tree-width.\ This is
possible because: (1) an MSO$_{2}$ property of a graph is nothing but an
MSO\ property of its incidence graph and (2) the clique-width of the incidence
graph of a graph is linearly bounded in terms of its tree-width.\ Our
constructions are actually implementable and usable.\ We detail concrete
constructions of automata in this perspective.

\textbf{keywords}: monadic second-order logic, edge quantification,
tree-width, clique-width, algorithmic meta-theorem, fly-automaton.

\end{abstract}

\section{Introduction}

Graphs are finite and, either directed or undirected. Our goal is to check
their \emph{monadic second-order} (\emph{MSO}) properties by using finite
automata running on the terms that denote input graphs, and to obtain in this
way \emph{fixed-parameter tractable} (\emph{FPT})\ algorithms whose parameters
are tree-width or clique-width.\ We want these automata to be constructable
automatically from logical formulas and practically usable.\ We recall the
following algorithmic meta-theorem \cite{CouEng, BCID12}\footnote{The
presentation in these two works is much better than those in the usually
quoted original references \cite{Cou90,CMR}.}; the notions it uses will be
reviewed in Section 2.

\bigskip

\textbf{Theorem 1 : \ }(a) For every integer $k$ and every MSO\ sentence
$\varphi$, there exists a linear time algorithm that checks the validity of
$\varphi$\ in any graph of clique-width at most $k$ given by a relevant term.

(b) For every integer $k$ and every MSO$_{2}$\ sentence $\varphi$, there
exists a linear time algorithm that checks the validity of $\varphi$\ in any
graph of tree-width at most $k$ given by a relevant tree-decomposition.\ 

\bigskip

A \emph{sentence} is a logical formula without free variables.\ An
MSO\ sentence can only use quantifications over vertices and sets of vertices,
whereas an MSO$_{2}$\ sentence can also use quantifications over edges and
sets of edges. The \emph{incidence graph} $Inc(G)$\ of a graph $G$\ is a
bipartite graph constructed as follows: its vertices are those of $G$ together
with new vertices representing its edges; its adjacency relation is the
incidence relation of $G$ (relating an edge and its ends). Assertion
(b)\footnote{By a result of Bodlaender (see \cite{Bod,DF,DF2}), a
tree-decomposition of $G $ of width $k$ can be computed in linear time if
there exists one.\ Hence the variant of (b) where a tree-decomposition is not
given but must be computed also holds, but this variant is not a consequence
of (a).\ Furthermore, the linear time decomposition algorithm is not
practically implementable.} is actually a consequence of (a) because:

(1) an MSO$_{2}$ property of a graph $G$\ is nothing but an MSO property of
its incidence graph $Inc(G)$\ and

(2) if $G$ has tree-width $k$, then $Inc(G)$ has clique-width at most $f(k)$
for some fixed linear function $f$,

(3) a tree-decomposition of $G$ of width $k$ can be converted in linear time
(for fixed $k$) into a clique-width term of width at most $f(k)$ that defines
$Inc(G)$.

Point (1) is just a matter of definitions.\ Point (2), in particular the fact
that $f$ is linear ($f(k)\leq2k+5$, see details in Section 3) together with
the linear time transformation of (3) make practically usable this reduction
of (b) to (a).

\bigskip

\emph{Automata constructions.}

The classical proof of Assertion (a) constructs from an MSO\ sentence
$\varphi$ and a positive integer $k$ a finite automaton $\mathcal{A}%
_{\varphi,k}$ that takes as input a term $t$ of width at most $k$ that denotes
an input graph assumed of clique-width $\leq k.$ This automaton recognizes the
terms that denote graphs satisfying $\varphi$. The construction is done by
induction on the structure of $\varphi$. First, one constructs automata for
the atomic formulas that are of only two kinds\footnote{Automata
$\mathcal{A}_{\varphi,k}$ can also be defined for formulas $\varphi$\ with
free variables.\ See Theorem 5.} : $X\subseteq Y$ (set inclusion) and
$edg(X,Y)$ (meaning that $X=\{x\},Y=\{y$\} and there is an edge from $x$ to
$y$, or between them if the graph is undirected).\ As we will recall,
MSO\ formulas can be written with set variables only and these two types of
atomic formulas.\ From automata $\mathcal{A}_{\varphi,k}$ and $\mathcal{A}%
_{\psi,k}$, one can build automata for $\mathcal{A}_{\varphi\vee\psi,k}$ and
$\mathcal{A}_{\varphi\wedge\psi,k}$ by taking the product of $\mathcal{A}%
_{\varphi,k}$ and $\mathcal{A}_{\psi,k}$ equipped with appropriate accepting
states. Other constructions handle negation and existential quantification.
See \cite{CouEng, BCID12}.

It is actually useful to precompute automata $\mathcal{A}_{\varphi,k}$ for
some formulas $\varphi$\ expressing basic MSO properties, such as
$Partition(X_{1},...,X_{p})$ (meaning that\ $X_{1},...,X_{p}$\ \ is a
partition of the vertex set), \emph{stability} of $X$ (denoted by $St(X)$ and
meaning that there are no edges between vertices in $X$, equivalently, that
$X$ is an independent set), connectedness of the considered graph, existence
of directed or undirected cycles, degree bounds etc.\ See \ \cite{BCID12}. The
logical expression of a property can be made simpler as we allow, in some
sense, more atomic formulas.\ For example, 3-colorability is expressed by :
$\exists X_{1},X_{2},X_{3}.(Partition(X_{1},X_{2},X_{3})\wedge St(X_{1})\wedge
St(X_{2})\wedge St(X_{3}))$.

However, even with this technique, this construction is intractable because
$\mathcal{A}_{\varphi,k}$ has in most cases, even for $k=2$, so many states
that it cannot be implemented in the classical way. This is not avoidable
\cite{FriGro}.\ The notion of a \emph{fly-automaton }(an FA in short) remedies
to this fact. Its states are not listed in huge tables.\ Although numerous,
they have a common syntactic structure and can be described by
words\footnote{FA can have infinitely many states: a state can record the
(unbounded) number of occurrences of a particular symbol. We can thus
construct fly-automata that check properties that are not MSO
expressible\ (for example that a graph is regular or can be partitioned into
$p$ disjoint regular graphs).\ These automata yield FPT or XP algorithms
\cite{DF,DF2} for clique-width as parameter. By equipping fly-automata with
output functions, we can make them compute values attached to graphs: for
example, assuming that the input graph is $s$-colorable, the minimum size of
$X_{1}$ in an $s$-coloring $(X_{1},\ldots,X_{s})$. We can compute the number
of $p$-vertex colorings, and also the number of so-called of \emph{acyclic}
$4$-colorings of Petersen's graph: 10800. The number of acyclic $3$-colorings
of McGee's graph is 57024.\ See \cite{BCID13}.}.\ The necessary
transitions\ are not stored in tables but computed by (small) programs.\ A
deterministic FA having 2$^{2^{10}}$\ states needs and computes only
100\ states on a term of size 100.\ The maximum size of a state (the number of
bits for encoding it) used on an input term is more important to bound the
computation time than the total number of states. Implementation of FA has
been tested in significant cases \cite{BCID12,BCID13}.

\bigskip

\emph{The present contribution }:

Our objective is to apply to Assertion (b) the tools developped in
\cite{BCID12,BCID13}. However, the reduction of (b) to (a) necessitates some
work on automata.\ For instance, the clique-width operations allow to write
terms that do not define incidence graphs.\ For example, if some vertex
intended to represent an edge has degree more than 2, the defined graph is not
an incidence graph.\ We will define an automaton $\mathcal{A}_{CT}$ to check
whether a given term is \emph{correct}, i.e, defines an incidence graph.

Furthermore, in MSO$_{2}$\ logic seen as MSO logic on incidence graphs, the
property $edg(X,Y)$ is no longer atomic.\ It is expressed by:

\begin{quote}
"there exists a vertex representing an edge that is adjacent to the unique
vertex of $X$ and to the unique vertex of $Y$",
\end{quote}

hence, the automaton for this formula is more complicated than the one for
$edg(X,Y)$ in the case of MSO logic because of the quantification on vertices
representing edges. We will detail the construction of such an FA for
$edg(X,Y)$ and of other FA for related properties such as domination. We will
also construct FA for checking Hamiltonicity, a graph property that is
MSO$_{2}$-expressible but not MSO-expressible.\ 

We consider \emph{directed graphs} because they are in general algorithmically
more difficult to handle than undirected ones.\ In the present setting they
are no more difficult to deal with, and it is easy to simplify our automata,
constructed for directed graphs so as they handle undirected graphs.

\bigskip

Our method, that consists in computing automata from logical formulas, yields
usable algorithms very quickly as it consists in combining predefined automata
by means of a few standard procedures operating on FA.\ It has been
implemented\footnote{The system AUTOGRAPH that builds and runs FA is written
in LISP \cite{BCID11} and http://dept-info.labri.u-bordeaux.fr/\symbol{126}%
idurand/autograph}.\ The produced algorithms follow the general scheme of
dynamic programming over graph decomposition, however, specialized algorithms
for particular problems designed with \emph{ad hoc} data structures can be
quicker. Our purpose is to obtain by meta-algorithmic tools, usable algorithms
with reasonable, but not necessarly optimal, time complexity.

Section 2\ reviews definitions and constructions of automata.\ Although this
article is a continuation of \cite{BCID12}, we recall most definitions to make
it readable independently.\ In Section 3, we adapt definitions to incidence
graphs.\ Our constructions of automata for the correctness of terms and for
properties based on adjacency are in Section 4.\ Other properties, in
particular Hamiltonicity, are discussed in Section 5.\ We review related work
in our conclusion.

\section{Definitions and background constructions.}

\bigskip

First, some notation : $\mathcal{P}(X)$ denotes the powerset of a set $X$,
$\mathcal{P}_{f}(X)$, the set of its finite subsets and $\mathcal{P}_{\leq
m}(X)$ the set of its subsets of cardinality at most $m$ and $[X\rightarrow
Y]_{f}$ the set of partial functions : $X\rightarrow Y$ having a finite domain.\ 

A \emph{(functional) signature} is a set $F$\ of function symbols, where $f\in
F$ has arity $\rho(f)\in\mathbb{N}$ and $T(F)$ is the set of (finite) terms
over $F$.\ The set of positions of a term $t$ is $Pos(t)$ and positions are
formally denoted by Dewey words.\ They are considered as the nodes of a
labelled tree, and the terminology of trees will be used for expressing
properties of positions in terms.\ The positions of $t=f(a,g(a,b))$ are the
words $\varepsilon$\ that it is the root of $t$, also denoted by $root_{t}$
and the unique occurrence of $f$, 1 and 21, the occurrences of $a$, 2 the
occurrence of $g$ and 22, the occurrence of $b$. If $t^{\prime}$ is the
subterm $t/u$ of $t$ issued from position $u$, then $Pos(t^{\prime})$ \ is the
set of words $w$ such that $uw\in Pos(t)$. Positions of $t$ are partially
ordered by $\leq_{t}$ \ such that $u\leq_{t}v$ if and only if $v$ is a prefix
of $u$, i.e., $v$ is an ancestor of $u$ or is equal to it.

By a \emph{language}, we mean a set of words or of terms, not to be confused
with a \emph{logical language} such as \emph{monadic second-order} (\emph{MSO}
in short) \ logic.

\bigskip

\emph{Graphs and MSO logic.}

\bigskip

A \emph{simple graph} is, here, a finite directed graph without parallel edges
and loops.\ These graphs will be used as \emph{incidence graphs} of more
general graphs, to be defined in the next section.\ We identify a simple graph
$G$\ to the logical structure $\langle V_{G},edg_{G}\rangle$ whose domain is
$V_{G}$, the set of vertices, and where $edg_{G}$ is the binary relation such
that $(x,y)\in edg_{G}$\ if and only if there is an edge from $x$ to $y$,
which we denote by $x\rightarrow_{G}y,$ or by $x\rightarrow y$ if $G$ is clear
from the context.

MSO\ logic uses individual variables ($x,y,z,...$) to denote vertices, and set
variables $(X,Y,Z,...)$ to denote set of vertices. Quantifications over binary
relations is not allowed.\ Rather than giving a formal syntax (see
\cite{CouEng}), we take the example of 3-vertex colorability, expressed by the
MSO\ formula $\exists X,Y.\alpha(X,Y)$ where $\alpha(X,Y)$ is the formula

\begin{quote}
$\qquad\qquad X\cap Y=\emptyset\wedge\forall u,v.\{edg(u,v)\Longrightarrow
\lbrack\lnot(u\in X\wedge v\in X)$

$\qquad\qquad\qquad\wedge\lnot(u\in Y\wedge v\in Y)\wedge\lnot(u\notin X\cup
Y\wedge v\notin X\cup Y)]\}.$
\end{quote}

The formula $\alpha(X,Y)$ expresses that $X,Y$ and $V_{G}-(X\cup Y)$ are the
three color classes of a 3-vertex coloring of $G$. More
generally,\textrm{\ \emph{p-}}vertex colorability\textrm{\ }is expressible in
a similar way.\ Other MSO\ expressible graph properties are connectedness,
strong connectedness, planarity (via forbidden minors) and the existence of
directed cycles.

Every MSO\ formula can be written so as to use only set variables and the
atomic formulas $X\subseteq Y$ (for set inclusion) and $edg(X,Y)$ (meaning
that $X=\{x\},Y=\{y$\} and $x\rightarrow y$): for doing that, we replace the
atomic formula $x\in Y$ by $X\subseteq Y$ where $X$ denotes a singleton set.
The property that $X$\ is singleton, denoted by $Sgl(X),$ can be expressed
itself in terms of inclusions.\ It is however useful to allow $X=Y,X=\emptyset
,$\ $Sgl(X)$, $Partition(X_{1},...,X_{p})$ (meaning that $(X_{1},...,X_{p})$
is a partition of the vertex set, where some sets $X_{i}$ may be empty) as
atomic formulas. Furthermore, universal quantifications $\forall X.\varphi
$\ are replaced by $\lnot\exists X.\lnot\varphi.$ These syntactic constraints
facilitate the construction of automata. Details are in \cite{CouEng}.

A \emph{sentence} is a formula without free variables.\ It expresses a
property of the considered graph.

$\bigskip$

$\emph{Clique-width}$

\bigskip

\emph{Clique-width}\ is a graph complexity measure, comparable to tree-width,
that is defined from operations that construct simple graphs equipped with
vertex labels. Let $C$ be a finite or countable set of labels. A
$C$-\emph{graph} is a triple $G=\langle V_{G},edg_{G},\pi_{G}\rangle$ where
$\pi_{G}$ is a mapping: $V_{G}\rightarrow C$. If $\pi_{G}(x)=a$ we say to be
short that $x$ is an $a$-\emph{vertex}.\ 

We let $F_{C}$ be the following set of operations on $C$-graphs:

\begin{quote}
$\oplus$ is the union of two disjoint $C$-graphs,

$relab_{a\rightarrow b}$, for $a\neq b$, $a,b\in C$, is the the unary
operation that changes every vertex label $a$ into $b$,

$\overrightarrow{add}_{a,b}$, for $a\neq b$, $a,b\in C$, is the unary
operation that adds an edge from each $a$-vertex $x$ to each $b$-vertex $y$
(unless we already have an edge from $x$ to $y$),

$\varnothing$ is a nullary symbol denoting the empty graph,

and for each $a\in C,$ the nullary symbol $\mathbf{a}$ denotes an isolated $a
$-vertex.
\end{quote}

\bigskip

The set $F_{C}$ is finite if $C$ is finite.

Every term $t$ in $T(F_{C})$ defines a $C$-graph $G=val(t)$ (a formal
definition is in \cite{BCID12, CouEng}). Its vertex set is the set of
positions in $t$ of the nullary symbols $\mathbf{a}$, $a\in C$. The
clique-width of a graph $G$, denoted by $cwd(G),$ is the least cardinality of
a set of labels $C$ such that $G$ is isomorphic, up to vertex labels, to
$val(t)$ for some $t$ in $T(F_{C})$. It is clear that $cwd(G)\leq\left\vert
V_{G}\right\vert .$

\bigskip

Let $t^{\prime}$ be a subterm $t/u$ of $t\in T(F_{C})$ for $u\in Pos(t).$ The
graph $val(t^{\prime})$ is, up to vertex labels, isomorphic to a subgraph
$G^{\prime}$ of $G=val(t).$ More precisely, let $i_{u}:Pos(t^{\prime
})\rightarrow Pos(t)$ map $w$ to $uw$ (we recall that positions are Dewey
words). Then $i_{u}(Pos(t^{\prime}))=\{v\in Pos(t)$ $\mid v\leq_{t}u\}$ is the
set of vertices of $G^{\prime}$ and $i_{u}$ is the isomorphism :
$val(t^{\prime})\rightarrow G^{\prime}$.\ If $x\rightarrow_{val(t^{\prime})}%
y$, then $i_{u}(x)\rightarrow_{G}i_{u}(y)$, however we may have $i_{u}%
(x)\rightarrow_{G}i_{u}(y)$ without having $x\rightarrow_{val(t^{\prime})}y$
because an edge from $i_{u}(x)$\ to $i_{u}(y)$ can be added to $G^{\prime}$ by
an operation $\overrightarrow{add}_{a,b}$\ above $u$ in $t$. To simplify
notation, we will use in the above case the following:

\begin{quote}
\emph{Convention about subterms}: we will forget $i_{u}$ and consider
$val(t^{\prime})$ as identical to the subgraph $G^{\prime}$ of $G=val(t)$.\ 
\end{quote}

Hence, a node $w$ of $val(t^{\prime})$ will be identified to the node
$i_{u}(w)=uw$ of $G$.

A term $t$ in $T(F_{C})$ is \emph{irredundant} if an operation
$\overrightarrow{add}_{a,b}$ never "tries to create" an edge from an
$a$-vertex $x$ to a\ $b$-vertex $y$ such that we already have $x\rightarrow
y$. In particular, $t$ has no subterm of the form $\overrightarrow{add}%
_{a,b}(t^{\prime})$ such that $val(t^{\prime})$ has an $a$-vertex $x$ and a
$b$-vertex $y$ such that $x\rightarrow_{val(t^{\prime})}y$.\ (The formal
definition is more complicated because of relabellings).\ Every term can be
made irredundant by an easy preprocessing \cite{BCID12, CouEng} consisting in
removing some operations $\overrightarrow{add}_{a,b}$.

\bigskip

\textbf{Lemma 2}\ : Let $t\in T(F_{C})$ be irredundant, $u\in Pos(t),t^{\prime
}=t/u.$ If a vertex $x$ of $val(t^{\prime})$ has indegree $p$ in this graph,
then it has indegree $p+q$ in $val(t)$, where $q$ is nonnegative and depends
only on $t,u$ and the label of $x$ in $val(t^{\prime}).$ The same holds for outdegrees.

\bigskip

This statement uses our \emph{Convention about subterms}.\ Its proof is
straightforward from the definition (Definition 5 of \cite{BCID12}).\ It
follows that if $t,t^{\prime},x,q$ are as in the statement and $x^{\prime}$ is
another vertex of $val(t^{\prime})$ with same label and of indegree
$p^{\prime}$, then it has indegree $p^{\prime}+q$ in $val(t)$.

\bigskip

We denote by $twd(G)$ the tree-width of a graph $G$ \cite{Bod,CouEng,DF,DF2}.

\bigskip

\textbf{Proposition 3} : For every simple graph $G$, we have $cwd(G)\leq
2^{2twd(G)+2}+1$, and $cwd(G)$ is not polynomially bounded in terms of
$twd(G)$, as it can be at least $2^{k-1}$ for some graphs of tree-width $2k$.

\bigskip

References for the proofs and comments are in \cite{CouEng}, Proposition 2.114.

\bigskip

Finally, we explain how a term in $T(F_{C})$\ can be enriched so as to define,
not only a graph, but also vertex sets $X_{1},...,X_{p}$ of this graph.\ We
replace in $F_{C}$\ each nullary symbol $\mathbf{a}$ by the nullary symbols
$(\mathbf{a},w)$ for all $w\in\{0,1\}^{p}.$ We obtain a set of operations
$F_{C}^{(p)}.$

Let $t\in T(F_{C})$, $X_{1},...,X_{p}\subseteq V_{val(t)}$ and $x\in
V_{val(t)}$.\ We let $w(x)$ be the sequence in $\{0,1\}^{p}$ whose $i$-th
element $w(x)[i]$ is 1 if $x\in X_{i}$ and 0\ otherwise. We replace in $t$ a
nullary symbol $\mathbf{a}$ at position $u$ by $(\mathbf{a},w(u))$ ($u$ is a
vertex of $val(t)$). We obtain a term\ in $T(F_{C}^{(p)})$ that defines the
graph $val(t)$ and also the $p$-tuple $(X_{1},...,X_{p})$. We denote this term
by $t\ast(X_{1},...,X_{p})$. Conversely, every term $t^{\prime}$ in
$T(F_{C}^{(p)})$ is of this form, and we will also denote by $val(t^{\prime})
$ the graph $val(t)$.\ 

If $P$ is a property of graphs\footnote{Saying that $P$ is a property of
graphs means that it is invariant under isomorphisms and arbitrary vertex
relabellings.\ Labels are used for constructing graphs, but at the end, they
are forgotten.}, then $L_{P,C}$ is defined as the set of terms $t\in T(F_{C}%
)$\ such that $val(t)$ satisfies $P$.\ More generally, if $P(X_{1},...,X_{p})$
is a property of vertex sets $X_{1},...,X_{p}$ of graphs, then $L_{P(X_{1}%
,...,X_{p}),C}$ is the set of terms

$t\ast(X_{1},...,X_{p})\in T(F_{C}^{(p)})$\ such that $P(X_{1},...,X_{p})$
holds in $val(t)$. \ If $P(X_{1},...,X_{p})$ is expressed by an
MSO\ formula\ $\varphi(X_{1},...,X_{p})$, we denote by $L_{\varphi
(X_{1},...,X_{p}),C}$ the language $L_{P(X_{1},...,X_{p}),C}$.

\bigskip

\emph{Automata.}

\bigskip

If $\varphi(X_{1},...,X_{p})$ is an MSO\ formula\ and $C$ is finite, then
$L_{\varphi(X_{1},...,X_{p}),C}$ is a recognizable language, hence, is
definable by a finite automaton. However, if $C$ is infinite, $L_{\varphi
(X_{1},...,X_{p}),C}$ can be defined by an infinite automaton that is usable
in algorithms.

\bigskip

\textbf{Definition 4: \ }\textit{Fly-automata.}

(a) A set is \emph{effectively given} if, either it is finite and the list of
its elements is known, or it is countably infinite and its elements are
bijectively encoded by words over $\{0,1\}$ (or another finite alphabet) that
form a decidable set. Through this encoding, we have the notion of
\emph{computable functions} on effectively given sets.\ If $\mathcal{D}$ is
effectively given, then so are $\mathcal{D}^{s}$ for $s\geq2$, $\mathcal{P}%
_{f}(\mathcal{D})$, the set of finite subsets of $\mathcal{D}$ and the set
$[\mathcal{D}^{\prime}\rightarrow\mathcal{D}]_{f}$ of partial functions having
a finite domain, where $\mathcal{D}^{\prime}$ is effectively given. A
signature $F$ is \emph{effectively given }if the set $F$ is effectively given,
the arity mapping $\rho$\ is computable and the least upper bound $\rho(F)$ of
the arities is finite.

\bigskip

(b) Let $F$ be an effectively given signature. A \emph{fly-automaton} over $F$
(in short, an FA\ over $F$) is a $4$-tuple $\mathcal{A}=\langle
F,Q_{\mathcal{A}},\delta_{\mathcal{A}},\mathit{Acc}_{\mathcal{A}}\rangle$ such
that $Q_{\mathcal{A}}$ is an effectively given set called the set of
\emph{states}, $\mathit{Acc}_{\mathcal{A}}$ is a decidable subset of
$Q_{\mathcal{A}}$ called the set of \emph{accepting states}%
\footnote{Fly-automata can also compute functions if one replaces the
accepting states by a computable mapping from states to some effective domain.
See \cite{BCID13}.}, and $\delta_{\mathcal{A}}$ is a computable function such
that, for each tuple $(f,q_{1},\dots,q_{\rho(f)})$ such that $q_{1}%
,\dots,q_{\rho(f)}\in Q_{\mathcal{A}}$ and $f\in F,$ the set of states
$\delta_{\mathcal{A}}(f,q_{1},\dots,q_{\rho(f)})$ \ is finite.\ The
\emph{transitions} are the pairs $f[q_{1},\dots,q_{\rho(f)}]\rightarrow
_{\mathcal{A}}q$ (and $f\rightarrow_{\mathcal{A}}q$ if $f$ is nullary) such
that $q\in\delta_{\mathcal{A}}(f,q_{1},\dots,q_{\rho(f)})$. We say that
$f[q_{1},\dots,q_{\rho(f)}]\rightarrow_{\mathcal{A}}q$ \ is a transition that
\emph{yields} $q$. We say that $\mathcal{A}$ is \emph{finite} if $F$ and
$Q_{\mathcal{A}}$ are finite, and we get the usual notion of a finite automaton.

\bigskip

(c) A \emph{run} of an FA $\mathcal{A}$ on a term $t\in T(F)$ is a mapping
$r:\mathit{Pos}(t)\rightarrow Q_{\mathcal{A}}$ such that :

\begin{quote}
if $u$ is an occurrence of a function symbol $f\in F$ and $u_{1}%
,...,u_{\rho(f)}$ is the sequence of sons of $u$, then $f[r(u_{1}%
),\dots,r(u_{\rho(f)})]\rightarrow_{\mathcal{A}}r(u)$; (if $\rho(f)=0$, the
condition reads $f\rightarrow_{\mathcal{A}}r(u)$).
\end{quote}

For $q\in Q_{\mathcal{A}}$, we define $L(\mathcal{A},q)$ as the set of terms
$t$ in $T(F)$ on which there is a run $r$ of $\mathcal{A}$ such that
$r(\mathit{root}_{t})=q$. A run $r$ on $t$ is \emph{accepting} if the state
$r(\mathit{root}_{t})$ is accepting. We define $L(\mathcal{A}):=\bigcup
\{L(\mathcal{A},q)\mid q\in\mathit{Acc}_{\mathcal{A}}\}\subseteq T(F)$. It is
the \emph{language accepted} (or \emph{recognized}) by $\mathcal{A}$. A state
$q$ is \emph{accessible} if $L(\mathcal{A},q)\neq\emptyset$. We denote by
$Q_{\mathcal{A}}\upharpoonright t$ the\ set of states that occur in the runs
on $t$ and on its subterms, and by $Q_{\mathcal{A}}\upharpoonright L$
the\ union of the sets $Q_{\mathcal{A}}\upharpoonright t$ for $t\in L\subseteq
T(F)$.

We denote by $\mathit{run}_{\mathcal{A},t}^{\ast}$ the mapping:
$Pos(t)\rightarrow\mathcal{P}_{f}(Q_{\mathcal{A}})$ that associates with every
position $u$ the \emph{finite set of states} of the form $r(root_{t/u}) $ for
some run $r$ on the subterm $t/u$ of $t$ issued from $u$. This mapping is
computable and the membership\ in $L(\mathcal{A})$ of a term in $T(F)$ is
decidable because $t\in L(\mathcal{A})$ if and only if the set\ $\mathit{run}%
_{\mathcal{A},t}^{\ast}(root_{t})$\ contains an accepting state. Hence,
whether all states of an FA\ are accessible or not does not affect the
membership algorithm: the inaccessible states simply never appear in any
run.\ There is no need to try to remove them, which is actually impossible in
general, unless if $\mathcal{A}$ is finite.\ Removing them in the case of a
"small" finite automaton permits to reduce the size of small the transition table.

We define $ndeg_{\mathcal{A}}(t)$, the \emph{nondeterminism degree of}
$\mathcal{A}$ \emph{on} $t,$ as the maximal cardinality of $\mathit{run}%
_{\mathcal{A},t}^{\ast}(u)$ for $u$ in $Pos(t)$. We have $ndeg_{\mathcal{A}%
}(t)\leq|Q_{\mathcal{A}}\upharpoonright t|.$\ 

A \emph{sink} is a state $s$ such that, for every transition $f[q_{1}%
,\dots,q_{m}]\rightarrow_{\mathcal{A}}q,$ we have $q=s$ if $q_{i}=s$ for some
$i$. If $F$ has at least one symbol of arity at least 2, then an automaton has
at most one sink. A state named $Error$ (resp.\ $Success$) will always be a
nonaccepting (resp.\ accepting) sink.

\bigskip

We say that an FA $\mathcal{B}=\langle H,Q_{\mathcal{B}},\delta_{\mathcal{B}%
},\mathit{Acc}_{\mathcal{B}}\rangle$ is a \emph{subautomaton} of an FA
$\mathcal{A}$ if $H\subseteq F$, $Q_{\mathcal{B}}\subseteq Q_{\mathcal{A}}$,
$\mathit{Acc}_{\mathcal{B}}=\mathit{Acc}_{\mathcal{A}}\cap Q_{\mathcal{B}}$
and $\delta_{\mathcal{B}}(f,q_{1},\dots,q_{\rho(f)})=\delta_{\mathcal{A}%
}(f,q_{1},\dots,q_{\rho(f)})$ if $f\in H$ and $q_{1},\dots,q_{\rho(f)}\in
Q_{\mathcal{B}}.$ Then $L(\mathcal{B})=L(\mathcal{A})\cap T(H).$

\bigskip

(d) \textit{Deterministic automata}. An FA $\mathcal{A}$ is
\emph{deterministic} (implicitly, \emph{and complete}) and if all sets
$\delta_{\mathcal{A}}(f,q_{1},\dots,$ $q_{\rho(f)})$\ have cardinality 1. A
deterministic FA $\mathcal{A}$ has, on each term $t$, a unique run denoted by
$run_{\mathcal{A},t};$ we let $q_{\mathcal{A}}(t):=run_{\mathcal{A}%
,t}(root_{t})$.\ The mapping $q_{\mathcal{A}}$\ is computable.

The computation time of $\mathcal{A}$ on a term $t$ is bounded by
$p.\left\vert t\right\vert $ where $p$ bounds the time used for computing a transition.\ 

\bigskip

For every fly-automaton $\mathcal{A}$, there exists a deteministic FA
$\mathcal{B}$ denoted by $\det(\mathcal{A})$ such that $Q_{\mathcal{B}%
}=\mathcal{P}_{f}(Q_{\mathcal{A}})$, $\mathit{run}_{\mathcal{B},t}%
=\mathit{run}_{\mathcal{A},t}^{\ast}$ and $L(\mathcal{B})=L(\mathcal{A})$.
(See \cite{BCID12}, Proposition 45(2)).\ A run of\ $\mathcal{B}$\ on a term
$t$, that is, a bottom-up computation of $\mathit{run}_{\mathcal{B},t}$, is
called \emph{the determinized run} of $\mathcal{A}$. The maximal size of a
state of $\det(\mathcal{A})$ on a term $t$ is bounded by $ndeg_{\mathcal{A}%
}(t).s$ where $s$ is the maximal size of a state in $Q_{\mathcal{A}%
}\upharpoonright t.$

If the state $Error$ is met at any point of the computation of a deterministic
FA, the term can be immediately rejected.\ If the state $Success$ is met at
any point of the bottom-up computation of an FA, then the term can be
immediately accepted. Hence, using sinks $Success$ and $Error$ \ in this way
can shorten some computations.

\bigskip

(e) \emph{Direct and inverse images of automata}.

Let $H$ and $F$\ be effectively given signatures. Let\ $h:H\rightarrow F$ be a
\emph{relabelling}, i.e., an arity preserving mapping having a
\emph{computable inverse}, that is, such that $h^{-1}(f)\subseteq H$ is finite
and computable for each $f$. Let $h$ be extended \ from $T(H)$ to $T(F)$\ in
the obvious way. If $L\subseteq T(H)$, then $h(L):=\{h(t)\mid t\in L\}$. If
$\mathcal{A}$ is an FA over $H$, then $h(\mathcal{A})$ is the fly-automaton
over $F$ (\cite{BCID12}, Proposition 45) obtained from $\mathcal{A}$ by
replacing each transition $f[q_{1},\cdots,q_{\rho(f)}]\rightarrow
_{\mathcal{A}}q$ by $h(f)[q_{1},\cdots,q_{\rho(f)}]\rightarrow q$. We say that
$h(\mathcal{A})$ is the \emph{image of} $\mathcal{A}$ under $h$. It is not
deterministic in general, even if $\mathcal{A}$ is. We have $h(L(\mathcal{A}%
))=L(h(\mathcal{A}))$ (because $h(\mathcal{A})$ has the same accepting states
as $A$).

Let now $h:T(H)\rightarrow T(F)$ be a computable relabelling. If $K\subseteq
T(F)$, then $h^{-1}(K):=\{t\in T(H)\mid h(t)\in K\}$. If $\mathcal{A}$ is an
FA over $F$, we let $h^{-1}(\mathcal{A})$ be the FA over $H$ with transitions
of the form $f[q_{1},\cdots,q_{\rho(f)}]\rightarrow q$ whenever $h(f)[q_{1}%
,\cdots,q_{\rho(f)}]\rightarrow_{\mathcal{A}}q$. We have $L(h^{-1}%
(\mathcal{A}))=h^{-1}(L(\mathcal{A}))$. We call $h^{-1}(\mathcal{A})$ the
\emph{inverse image} of $\mathcal{A}$ under $h$ (\cite{BCID12}, Definition
17(h)). It is deterministic if $\mathcal{A}$ is so.

\bigskip

\emph{Automata from checking graph properties.}

\bigskip

\textbf{Theorem 5 }: Let $\varphi(X_{1},...,X_{p})$ be an MSO\ formula and $C$
be an effectively given set of labels.\ 

(1) There exists a deterministic FA $\mathcal{A}_{\varphi(X_{1},...,X_{p}),C}$
over $F_{C}$ that recognizes the language\ $L_{\varphi(X_{1},...,X_{p}),C}%
$.\ This automaton can be constructed by an algorithm.

(2) If $C^{\prime}\subseteq C$\ is effectively given, then $\mathcal{A}%
_{\varphi(X_{1},...,X_{p}),C^{\prime}}$ is a subautomaton of $\mathcal{A}%
_{\varphi(X_{1},...,X_{p}),C}$.\ 

(3) If $C$ is finite, then $\mathcal{A}_{\varphi(X_{1},...,X_{p}),C}$ is finite.

\bigskip

\textbf{Proof} : In \cite{CouEng, BCID12} and related articles, we construct,
for each finite set $C$, a finite deterministic automaton $\mathcal{A}%
_{\varphi(X_{1},...,X_{p}),C}$ that accepts the recognizable language
$L_{\varphi(X_{1},...,X_{p}),C}\subseteq T(F_{C}^{(p)})$.\ The construction is
by induction on the structure of $\varphi(X_{1},...,X_{p})$.\ We first
construct automata for the atomic formulas, $X\subseteq Y$ and $edg(X,Y)$ and
also for some useful MSO properties such as $X=\emptyset$, $Sgl(X)$,
$Partition(X_{1},\ldots,X_{p})$, $Conn(X)$ expressing that the induced
subgraph with vertex set $X$\ is connected, $DirCycle$ expressing that the
graph has an directed cycle.\ These constructions are based on our
understanding of the considered properties.

Operations on FA "implement" the logical connectives $\vee,\wedge
,\lnot,\exists$ and variable substitutions.\ The operation reflecting
existential quantification consists in taking an image automaton, followed by
a determinization.\ We have :

\begin{quote}
$\mathcal{A}_{\exists X_{p}.\varphi(X_{1},...,X_{p}),C}=\det(h(\mathcal{A}%
_{\varphi(X_{1},...,X_{p}),C}))$\ 
\end{quote}

where $h:F_{C}^{(p)}\rightarrow F_{C}^{(p-1)}$ deletes the last Boolean of $w
$ in each nullary symbol $(\mathbf{a},w)$.\ 

The transformation reflecting substitutions of variables consists in taking an
inverse image: if $\psi(X_{1},...,X_{p})$ is defined as $\varphi(X_{i_{1}%
},...,X_{i_{m}})$ from $\varphi(Y_{1},...,Y_{m})$, then the automaton
$\mathcal{A}_{\psi(X_{1},...,X_{p}),C}$ is an inverse image of $\mathcal{A}%
_{\varphi(Y_{1},...,Y_{m}),C}$\ (\cite{BCID12}, Lemma 13 and Section
4.2.1).\ For an example, if $\psi(X_{1},X_{2},X_{3})$\ is defined as
$edg(X_{3},X_{1}),$ we obtain $\mathcal{A}_{\psi(X_{1},X_{2},X_{3}),C}%
=h^{-1}(\mathcal{A}_{edg(X_{1},X_{2}),C})$ where $h((\mathbf{a},w_{1}%
w_{2}w_{3}))$ \ $=(\mathbf{a},w_{3},w_{1})$\ for all $w_{1},w_{2},w_{3}%
\in\{0,1\}$. If $\varphi(X_{1},...,X_{p})$ is the conjunction of two formulas
for which we know automata, then the automaton for $\varphi(X_{1},...,X_{p})$
is the product of these two automata, provided they are built for formulas
having the same list of free variables.\ Variable substitution is useful to
insure this technical point\footnote{We use this observation in Section 4.2.}.\ 

\bigskip

As explained in \cite{BCID12}, Section 7.3.1, these constructions work for
infinite sets of labels. The three assertions of Theorem 5 can be proved by
induction on the structure of\ $\varphi$.\ To be more precise\footnote{This
definition is used in \cite{BCID14} to prove that the\emph{\ (Strong)
Recognizability Theorem, }Theorem 5.68(1) of \cite{CouEng}, can be proved via
the construction of automata.}, we define, for each set $X$\ and integer
$i\geq0$, the set $\mathcal{L}_{i}(X)$ as follows:

\begin{quote}
$\mathcal{L}_{0}(X):=X,$

$\mathcal{L}_{i+1}(X):=\mathcal{L}_{i}(X)\cup\mathcal{P}_{f}(\mathcal{L}%
_{i}(X))\cup(\mathcal{L}_{i}(X)\times\mathcal{L}_{i}(X)).$
\end{quote}

It is clear that $X\subseteq Y$ implies $\mathcal{L}_{i}(X)\subseteq
\mathcal{L}_{i}(Y)$ and that $\mathcal{L}_{i}(X)$ is finite if $X$ is finite.
The set $\mathcal{L}_{i}(X)$ is effectively given if $X$ is.\ 

For every MSO formula $\varphi(X_{1},...,X_{p})$, one can define a finite set
$B$ and an integer $i$ such that, for each effectively given set $C$ of
labels, one can construct a deterministic FA $\mathcal{A}_{\varphi
(X_{1},...,X_{p}),C}$ over $F_{C}^{(p)}$ that recognizes the language
$L_{\varphi(X_{1},...,X_{p}),C}$ and whose set of states $Q(C)$ satisfies the
following properties:

\begin{quote}
(i) $Q(C)\subseteq\mathcal{L}_{i}(B\cup C),$

(ii) $Q(C^{\prime})=Q(C)\cap\mathcal{L}_{i}(B\cup C^{\prime})$ if $C^{\prime
}\subseteq C,$

(iii) if $\ell:C\rightarrow C^{\prime}$ is a bijection, then $\mathcal{A}%
_{\varphi(X_{1},...,X_{p}),C^{\prime}}$ is the FA obtained from $\mathcal{A}%
_{\varphi(X_{1},...,X_{p}),C}$ by replacing in its states and transitions $a$
by $\ell(a)$ for each $a\in C$ (informally, $\ell$ yields an isomorphism of
automata : $\mathcal{A}_{\varphi(X_{1},...,X_{p}),C}\rightarrow\mathcal{A}%
_{\varphi(X_{1},...,X_{p}),C^{\prime}}).$
\end{quote}

\bigskip

For an example, the automaton $\mathcal{A}_{edg(X,Y),C}$ \ constructed in
\cite{CouEng, BCID12} has states in $\{Error,Ok\}\cup(P_{\leq1}(C)\times
P_{\leq1}(C))$ that is a subset of $\mathcal{L}_{2}(\{Error,Ok\}\cup C).$ (As
they are defined in \cite{CouEng, BCID12}, states are in $\mathcal{L}%
_{1}(B\cup C)$ for some finite set $B$.)

\bigskip

It follows from Properties (i)-(iii) that we need only construct a single FA
$\mathcal{A}_{\varphi(X_{1},...,X_{p}),C}$ over $F_{C}^{(p)}$ where $C$ is an
infinite effectively given set of labels. We will take $C=\mathbb{N}_{+}$.
Running $\mathcal{A}_{\varphi(X_{1},...,X_{p}),C}$ on terms in $T(F_{[k]}%
^{(p)})$, where $[k]:=\{1,...,k\},$\ is the same as running $\mathcal{A}%
_{\varphi(X_{1},...,X_{p}),[k]}$.\ An implementation of $\mathcal{A}%
_{\varphi(X_{1},...,X_{p}),C}$ yields a single program that works for all
terms in $T(F_{[k]}^{(p)})$, $k\geq1$, hence for graphs of all clique-widths.

The membership in $L_{\varphi(X_{1},...,X_{p}),C}$ of a term $t\in
T(F_{[k]}^{(p)})$ can be decided in (FPT) time $f(k).\left\vert t\right\vert $
for a function $f$ depending only on $\varphi(X_{1},...,X_{p}).$ This function
bounds the time for computing transitions.\ This gives a linear time
recognition algorithm for graphs of clique-width at most some fixed $k$ given
by terms in $T(F_{[k]})$, but finding such terms takes (presently) cubic
time.\ See \cite{CouEng} for details and references.

\bigskip

\emph{Properties of induced subgraphs}

\bigskip

We apply to the \emph{relativization} of a property $P$ to a set $X$\ the
notion of inverse image of an FA. If $P$ is a graph property, we denote by
$P[X]$ the property of $(G,X)$ where $X$ is a set of vertices of $G$ such that
the induced subgraph $G[X]$ satisfies $P$.

Let $t\in T(F_{C})$ define a graph $G:=val(t)$ and $X$ be a set of vertices of
$G$. (We recall that $X$ is a set of occurrences of the nullary symbols
$\mathbf{a}$ for $a\in C$.) Let $t^{\prime}\in T(F_{C})$ be obtained from $t$
by replacing each $\mathbf{a}$ \ by the nullary symbol $\varnothing$\ (that
denotes the empty graph) at its occurrences not in $X$. Then $val(t^{\prime})$
is $G[X]$, the induced subgraph of $G$ with vertex set $X$. (If $t$ is
irredundant, then so is $t^{\prime}).$

\ Then the subset $L_{P[X],C}$ of $T(F_{C}^{(1)})$ is $h^{-1}(L_{P,C})$ where
$h:T(F_{C}^{(1)})\rightarrow T(F_{C})$\ replaces in a term of $T(F_{C}^{(1)})$
each nullary symbol $(\mathbf{a},0)$ by $\varnothing$\ and $(\mathbf{a},1)$ by
$\mathbf{a}$. Hence, if $P$ is defined by an MSO sentence $\varphi,$ then
$P[X]$ is defined by an MSO formula $\varphi\lbrack X]$ with free variable
$X$\ (we need not write this formula) and $\mathcal{A}_{\varphi\lbrack X],C}$
is the inverse image by $h$ of $\mathcal{A}_{\varphi,C}.\ $See Definition 4(e)
and Lemma 15\ of \cite{BCID12}, or \cite{CouEng}.

\bigskip

\emph{Irredundant terms}

\bigskip

The set of irredundant terms in $T(F_{C})$ or in $T(F_{C}^{(p)})$, denoted by
$L_{Irr}$, is recognized by an FA\ $\mathcal{I}$ (\cite{BCID12}), whose states
are $Error$ and the finite subsets of $C^{2}$.\ For each term $t$,
$q_{\mathcal{I}}(t)=Error$ if $t$ is not irredundant and is $\{(\pi
_{val(t)}(x),\pi_{val(t)}(y))\mid x\rightarrow_{val(t)}y\}$ if $t$ is
irredundant.\ Its transitions are easy to define.

We will construct automata $\mathcal{A}_{\varphi(X_{1},...,X_{p}),C}$ intended
to work correctly for irredundant terms.\ This means that $L(\mathcal{A}%
_{\varphi(X_{1},...,X_{p}),C})\cap L_{Irr}=L_{\varphi(X_{1},...,X_{p}),C}\cap
L_{Irr}.$ If an input term is not irredundant, the answer of $\mathcal{A}%
_{\varphi(X_{1},...,X_{p}),C}$ may be incorrect.\ By taking the product of
$\mathcal{A}_{\varphi(X_{1},...,X_{p}),C}$ and $\mathcal{I}$, we can get an
automaton that recognizes exactly $L_{\varphi(X_{1},...,X_{p}),C}\cap
L_{Irr}.$ However, we will assume that the verification of irredundancy is
done by the \emph{parsing algorithm} that builds the term denoting the input
graph to be processed by the automaton.

\section{Incidence graphs}

When the domain of the relational structure representing a graph is its vertex
set, and since quantifications over binary relations are not allowed, MSO
formulas cannot use quantified variables denoting sets of edges.\ In the
incidence graph $Inc(G)$ of a graph $G$, each edge of $G$\ is made into a
vertex, hence an MSO\ formula over $Inc(G)$ can be seen as an MSO formula
using quantifications on sets of edges and it is called an MSO$_{2}$ formula
intended to express a property of $G$.\ The formula : "there exists a set of
edges that induces a directed cycle and goes through all vertices", expresses
that the considered graph has a directed Hamiltonian cycle.\ This property is
not expressible in MSO\ logic over the relational $\langle V_{G}%
,edg_{G}\rangle$ \cite{CouEng}.

Assertion (b) of Theorem 1 does not extend to inputs $Inc(G)$\ for $G$\ of
clique-width at most $k$ because the clique-width of such graphs $Inc(G)$ is
unbounded. It can be used for graphs $G$\ of tree-width at most $k$%
.\ Proposition 3\ gives $cwd(Inc(G))=2^{O(twd(G))}$ (because $twd(Inc(G))\leq
twd(G)+1$) however $cwd(Inc(G))=O(twd(G))$ as we will see.\ 

\bigskip

\textbf{Definition 6 }\emph{: Incidence graphs.}

\bigskip

Let $G$ be a directed graph that can have parallel edges and loops. We define
it as a triple\ $\langle V_{G},E_{G},inc_{G}\rangle$ such that $V_{G} $\ is
its set of vertices, $E_{G}$\ is its set of edges (of course $V_{G}\ \cap
E_{G}=\emptyset$) and $inc_{G}\subseteq(V_{G}\times E_{G})\cup(E_{G}\times
V_{G})$ is such that, for $u,v\in V_{G}$ and $e\in E_{G}$, we have $(u,e)\in
inc_{G}$ and $(e,v)\in inc_{G}$ if $e$ is an edge $u\rightarrow v$. We define
$Inc(G)$ as the simple, directed and bipartite graph\ identified to the
relational structure\ $\langle V_{G}\cup E_{G},inc_{G},isv_{G}\rangle$ where
$isv$ is unary and $isv_{G}(u)$ holds if and only if $u\in V_{G}$. This
relation is useful to distinguish the vertices from the edges in an arbitrary
relational structure\ $\langle X,inc,isv\rangle$ isomorphic to\ $\langle
V_{G}\cup E_{G},inc_{G},isv_{G}\rangle$.\ It follows that $G$ can be
reconstructed up to isomorphism from any structure $\langle X,inc,isv\rangle$
isomorphic to\ $\langle V_{G}\cup E_{G},inc_{G},isv_{G}\rangle$.

If $G$ is undirected, then we define $inc_{G}\subseteq E_{G}\times V_{G}$
where $(e,u)\in inc_{G}$ if $u$ is an end of $e$. In this case, the predicate
$isv_{G}$ is not necessary because the vertices of $G$ are the elements $x$ of
the domain such that $inc(x,y)$ holds for no $y$. We will only discuss
directed graphs in the sequel, leaving to the reader the easy task of
simplifying our constructions so as to handle undirected graphs.

\bigskip

We will use some adaptations of MSO\ formulas and clique-width operations. We
will consider a relational structure $S=\langle D_{S},inc_{S},isv_{S}\rangle$
as a simple graph, also denoted by $S$, whose vertex set is $D_{S}$, adjacency
relation\ is $inc_{S}$ and that is equipped with a distinguished set of
vertices $isv_{S}$.\ In such a structure isomorphic to an incidence graph
$\langle V_{G}\cup E_{G},inc_{G},isv_{G}\rangle$, we will call $v$%
-\emph{vertices} the vertices of $S$ that satisfy $isv$ and represent the
vertices of $G$\ and $e$-\emph{vertices} the others, that represent its
edges.\ It is FO\ expressible whether a relational structure $S=\langle
D_{S},inc_{S},isv_{S}\rangle$\ is $Inc(G)$ for some directed graph $G$.\ If
this is true, then $G$ is unique and its vertex set is $isv_{G}$.

For expressing properties of such structures $S$, we will write MSO\ formulas
\ with two types of set variables : $X,Y,...$\ to denote sets of $v$-vertices
and $U,V,W$ to denote sets of $e$-\emph{vertices}\footnote{These formulas
could be alternatively defined as MSO\ formulas where each variable
$X$\ (resp. $U$) comes with the condition that the elements $x$ of $X$
(resp.\ of $U$) satisfy $isv(x)$ (resp.\ $\lnot isv(x)$).}. The atomic
formulas will be of the forms $X\subseteq Y,U\subseteq V,inc(X,U)$ and
$inc(U,X)$; a formula $X\subseteq U$ or $inc(X,Y)$ is not allowed, and the
predicate $isv$ will not occur in formulas, as it is replaced by the typing of
variables, forcing them to denote either sets of vertices or sets of edges.

\bigskip

For constructing incidence graphs with clique-width operations, we will use
pairs $(C,D)$ of disjoint, effectively given sets of labels. Those in $C$ will
define the $v$-vertices, and those in $D$ the $e$-vertices. The operations
$relab$ and $\overrightarrow{add}$ will be those from $F_{C\cup D} $\ such that:

\begin{quote}
no label in $C$ can be changed to a label in $D$, and vice-versa,

no edge-addition $\overrightarrow{add}_{a,b}$ can be used with $a,b\in C$ or
$a,b\in D$.
\end{quote}

We obtain an effectively given signature $F_{C,D}$. \ Every term $t\in
T(F_{C,D})$ defines a simple bipartite graph $val(t)=H=\langle V_{H}%
,inc_{H},isv_{H}\rangle$ where $isv_{H}(x)$ holds if and only if $x$ is
defined by $\mathbf{a}$ for some $a\in C$.\ We have $inc_{H}(x,y)$ if and only
if $x\rightarrow_{val(t)}y.$ We say that $t$ is \emph{correct} if $H=Inc(G)$
for some graph $G$, whose vertex set is then necessarly $isv_{H}$. This is the
case\ if and only if each vertex having a label in $D$ has indegree and
outdegree 1.

\bigskip

\textbf{Proposition 7\ }: If a directed graph $G$ has tree-width $k$, then
$Inc(G)$ is defined by a term in $T(F_{C,D})$ such that $\left\vert
C\right\vert =2$ and $\left\vert D\right\vert =2k+3.$ If $G$ is
undirected,\ then $Inc(G)$ is defined by a term in $T(F_{C,D})$ such that
$\left\vert C\right\vert =2$ and $\left\vert D\right\vert =k+2.$ Terms
witnessing these bounds can be constructed in linear time from
tree-decompositions of $G$ of width $k.$ Conversely, $twd(G)=O(cwd(Inc(G)))$.

\bigskip

\textbf{Proof }: See \cite{Bou,Cou15} for the first three assertions.\ We have
actually $cwd(Inc(G))$ $\leq2k+4$\ for $G$ directed if we allow relabellings
$a\rightarrow d$ such that $a\in C$ and $d\in D$\ in terms defining
$Inc(G)$.\ Similarly, $cwd(Inc(G))\leq k+3$ if $G$ is undirected.

The last assertion holds because $twd(G)\leq twd(Inc(G))$ and, by \cite{GW}
(also in \cite{CouEng}, Proposition 2.115),\ since $Inc(G)$ has no subgraph
isomorphic to $K_{3,3}$, we have $twd(Inc(G))\leq6.cwd(Inc(G))-1$.$\square$

\bigskip

By the last assertion, the reduction of (b) to (a) in Theorem 1 does not work
for the verification of MSO$_{2}$ properties of graphs of bounded
clique-width. \ This is not a surprise because there are MSO$_{2}$ properties
that are not FPT\ for clique-width unless \textsc{FPT = W[1]} (\cite{Fom})
which is unlikely, similarly to \textsc{N = NP} (see \cite{DF,DF2} for the
classes \textsc{FPT} and \textsc{W[1]}).

\bigskip

In order to represent in terms sets of vertices $X_{1},...,X_{p}$ \ and sets
of edges $U_{1},...,U_{m}$, we replace in $F_{C,D}$ each nullary symbol
$\mathbf{a}$, for $a\in C$, by the nullary symbols $(\mathbf{a},w)$ for
$w\in\{0,1\}^{p}$ and each nullary symbol $\mathbf{d}$, for $d\in D$, by the
symbols $(\mathbf{d},w)$ for $w\in\{0,1\}^{m}$.\ We obtain a signature
$F_{C,D}^{(p,m)}$ and, for each MSO\ formula $\varphi(X_{1},...,X_{p}%
,U_{1},...,U_{m}),$ a representing language $L_{\varphi(X_{1},...,X_{p}%
,U_{1},...,U_{m}),C,D}\subseteq T(F_{C,D}^{(p,m)})$ consisting of the correct
terms $t$ that encode an assignment to $X_{1},...,X_{p},$ $U_{1},...,U_{m}%
$\ for which $\varphi$\ holds in $val(t)$. We will construct FA that recognize
these languages.

\bigskip

\emph{Denotation of subgraphs}

\bigskip

Let $G$ be a graph, $X\subseteq V_{G}$ and $U\subseteq E_{G}$.\ We let
$Inc(G)[X,U]:=\langle X\cup U,inc_{G}\cap(X\cup U)^{2},isv_{G}\cap(X\cup
U)\rangle$. This structure is an incidence graph $Inc(H)$ if and only if the
ends of all "edges" $u\in U$ are in $X$.\ If this is the case, then $V_{H}=X$,
$E_{H}=U$ and $H$ is a subgraph of $G$. We call $Subgraph(X,U)$ this property
of $(G,X,U)$. Assume now that $Inc(G)=val(t)$ where $t\in T(F_{C,D})$, and
$X,U$ as above are sets of occurrences in $t$ of nullary symbols respectively
in $\mathbf{C}$ and in $\mathbf{D}$. Let $t[X,U]$ be obtained from $t$ by
replacing by $\varnothing$\ (denoting the empty graph) the nullary symbols at
their occurrences not in $X\cup U$. It is clear that
$val(t[X,U])=Inc(G)[X,U].$

Then $t[X,U]$ is a correct term if and only if $Inc(G)[X,U]$ is an incidence
graph. The set of correct terms $t[X,U]\in T(F_{C,D}^{(1,1)})$ is
\ $h^{-1}(L)$ where $L$ is the set of correct terms in $T(F_{C})$ and $h$ maps
$(\mathbf{a},1)$ to $\mathbf{a}$ and $(\mathbf{a},0)$ to $\varnothing$ for
$a\in C\cup D$.\textbf{\ }We will define $L$ by an FA $\mathcal{A}_{CT}$.\ The
property $Subgraph(X,U)$ will thus be defined by its inverse image,
$h^{-1}(\mathcal{A}_{CT})$.\ 

\section{Automata}

In this section, we construct FA to check MSO$_{2}$ properties of directed
graphs.\ It is easy to modify them in order to check similar properties of
undirected graphs.\ These automata will be deterministic and designed so as to
work correctly on irredundant terms in $T(F_{C,D})$ for pairs $(C,D)$ of
disjoint effectively given sets of labels.\ 

They will be \emph{linear FPT-FA}, meaning that their computation times are
linear in the size of input terms over fixed finite subsignatures of $F_{C,D}$.

Their constructions are the same for $C\cup D$ either finite or infinite, as
explained after Theorem 5. We will construct FA\ for unspecified pairs $C,D$,
either $C=\mathbb{N}_{+}$ and $D=\{-n\mid n\in\mathbb{N}_{+}\}$ (for being
concrete) or finite subsets of them.\ The complexities of our automata will
appear from their numbers of states when $C$ and $D$ are finite. An
FA\ constructed from a formula $\varphi$\ will be denoted by $\mathcal{A}%
_{\varphi}$ without reference to $C,D$.

\subsection{Correct terms}

Every term $t\in T(F_{C,D})$\ defines a simple bipartite graph $val(t)$, and
is correct if and only if $val(t)$ is an incidence graph.\ A $C$-vertex of
$val(t)$ is a vertex having a label in $C$, a $D$-vertex is one having a label
in $D$.\ These definitions apply even if $t$ is not correct.

We describe an FA\ $\mathcal{A}_{CT}$\ that checks the correctness of terms in
$T(F_{C,D})$, assumed to be irredundant.\ Its states are the sink $Error$ and
the 6-tuples $(\gamma_{1},\gamma_{2},\delta_{00},\delta_{01},\delta
_{10},\delta_{11})\in\mathcal{P}_{f}(C)^{2}\times\mathcal{P}_{f}(D)^{4}$\ such
that $\gamma_{1}\cap\gamma_{2}=\emptyset$ and $\delta_{00},\delta_{01},$
$\delta_{10},\delta_{11}$\ are pairwise disjoint.\ We will denote
$(\delta_{00},\delta_{01},\delta_{10},\delta_{11})$ by$\ \overrightarrow
{\delta}$.

At the root of a term $t\in T(F_{C,D}),$ $\mathcal{A}_{CT}$ reaches the state
$(\gamma_{1},\gamma_{2},\overrightarrow{\delta})$ if and only if, we have in
$val(t)$:

\begin{quote}
(i) $\gamma_{1}$ is the set of labels $a\in C$ that label a single $C$-vertex.

(ii) $\gamma_{2}$ is the set of labels $a\in C$ that label at least two $C$-vertices,

(iii) $\overrightarrow{\delta}=(\delta_{00},\delta_{01},\delta_{10}%
,\delta_{11})$ where for $i,j\in\{0,1\}$, $\delta_{ij}$ is the set of labels
of $D$-vertices of indegree $i$ and outdegree $j$,

(iv) the sets $\delta_{00},\delta_{01},\delta_{10},\delta_{11}$ defined in
(iii) are pairwise disjoint,

(v) no $D$-vertex has indegree or outdegree 2 or more.
\end{quote}

It reaches the state $Error$ if (iv) or (v) does not hold.

The accepting states are the tuples $(\gamma_{1},\gamma_{2},\overrightarrow
{\delta})$ such that $\delta_{00}=\delta_{01}=\delta_{10}=\emptyset.$\ 

For any subterm $t$ of a correct term, Condition (iv) is necessary by Lemma 2
and Condition (v) also, because the $D$-vertices in an incidence graph
represent edges. Transitions are listed in Table 1.\ In order to simplify the
table, we use the following notations and conventions. First, we do not list
the transitions with $Error$ on the left side as they always yield $Error$
(cf.\ Definition 4(c)). Furthermore,

\begin{quote}
$a,b$ denote elements of $C$ and $d,e$ denote elements of $D$,

$d\in\overrightarrow{\delta}$ means $d\in\delta_{00}\cup\delta_{01}\cup
\delta_{10}\cup\delta_{11},$

$Disj(\overrightarrow{\delta})$ means that $\delta_{00},\delta_{01}%
,\delta_{10},\delta_{11}$ are pairwise disjoint,

$\overrightarrow{\delta}\cup\overrightarrow{\delta^{\prime}}:=(\delta_{00}%
\cup\delta_{00}^{\prime},\delta_{01}\cup\delta_{01}^{\prime},\delta_{10}%
\cup\delta_{10}^{\prime},\delta_{11}\cup\delta_{11}^{\prime}),$

$\gamma\lbrack a\rightarrow b]$ is the set $\gamma$\ where $a$ is replaced by
$b$, and similarly for $\delta\lbrack d\rightarrow e],$

$\overrightarrow{\delta}[d\rightarrow e]:=(\delta_{00}[d\rightarrow
e],\delta_{01}[d\rightarrow e],\delta_{10}[d\rightarrow e],\delta
_{11}[d\rightarrow e]),$

$\overrightarrow{\emptyset}$ denotes a sequence of empty sets of appropriate length.
\end{quote}

\bigskip%
\begin{tabular}
[c]{|c|c|}\hline
Transitions & Conditions\\\hline\hline
\multicolumn{1}{|l|}{$\varnothing\rightarrow\overrightarrow{\emptyset}$} &
\multicolumn{1}{|l|}{}\\\hline
\multicolumn{1}{|l|}{$\mathbf{a}\rightarrow(\{a\},\overrightarrow{\emptyset})$
\ } & \multicolumn{1}{|l|}{}\\\hline
\multicolumn{1}{|l|}{$\mathbf{d}\rightarrow(\emptyset,\emptyset
,\{d\},\overrightarrow{\emptyset})$} & \multicolumn{1}{|l|}{}\\\hline
\multicolumn{1}{|l|}{$relab_{a\rightarrow b}[(\gamma_{1},\gamma_{2}%
,\overrightarrow{\delta})]\rightarrow(\gamma_{1},\gamma_{2},\overrightarrow
{\delta})$} & \multicolumn{1}{|l|}{$a\notin\gamma_{1}\cup\gamma_{2}$}\\\hline
\multicolumn{1}{|l|}{$relab_{a\rightarrow b}[(\gamma_{1},\gamma_{2}%
,\overrightarrow{\delta})]\rightarrow(\overline{\gamma}_{1},\overline{\gamma
}_{2},\overrightarrow{\delta})$} & \multicolumn{1}{|l|}{$\{a,b\}\subseteq
\gamma_{1}\cup\gamma_{2}$, $\overline{\gamma}_{1}=\gamma_{1}-\{a,b\}$}\\
& \multicolumn{1}{|l|}{and $\overline{\gamma}_{2}=\gamma_{2}\cup\{b\}-\{a\}$%
}\\\hline
\multicolumn{1}{|l|}{$relab_{a\rightarrow b}[(\gamma_{1},\gamma_{2}%
,\overrightarrow{\delta})]\rightarrow(\overline{\gamma}_{1},\overline{\gamma
}_{2},\overrightarrow{\delta})$} & \multicolumn{1}{|l|}{$a\in\gamma_{1}%
\cup\gamma_{2}$, $b\notin\gamma_{1}\cup\gamma_{2}$,}\\
\multicolumn{1}{|l|}{} & \multicolumn{1}{|l|}{$\overline{\gamma}_{1}%
=\gamma_{1}[a\rightarrow b]$ and $\overline{\gamma}_{2}=\gamma_{2}%
[a\rightarrow b] $}\\\hline
\multicolumn{1}{|l|}{$relab_{d\rightarrow e}[(\gamma_{1},\gamma_{2}%
,\overrightarrow{\delta})]\rightarrow(\gamma_{1},\gamma_{2},\overrightarrow
{\delta}[d\rightarrow e])$} & \multicolumn{1}{|l|}{$Disj(\overrightarrow
{\delta}[d\rightarrow e])$}\\\hline
\multicolumn{1}{|l|}{$relab_{d\rightarrow e}[(\gamma_{1},\gamma_{2}%
,\overrightarrow{\delta})]\rightarrow Error$} & \multicolumn{1}{|l|}{$\lnot
Disj(\overrightarrow{\delta}[d\rightarrow e])$}\\\hline
\multicolumn{1}{|l|}{$\overrightarrow{add}_{a,d}[(\gamma_{1},\gamma
_{2},\overrightarrow{\delta})]\rightarrow(\gamma_{1},\gamma_{2}%
,\overrightarrow{\delta})$} & \multicolumn{1}{|l|}{$a\notin\gamma_{1}%
\cup\gamma_{2}$ or $d\notin\overrightarrow{\delta}$}\\\hline
\multicolumn{1}{|l|}{$\overrightarrow{add}_{a,d}[(\gamma_{1},\gamma
_{2},\overrightarrow{\delta})]\rightarrow Error$} &
\multicolumn{1}{|l|}{($a\in\gamma_{2}$ and $d\in\overrightarrow{\delta}$)
or}\\
& \multicolumn{1}{|l|}{$a\in\gamma_{1}$ and $d\in\delta_{10}\cup\delta_{11}$%
}\\\hline
\multicolumn{1}{|l|}{$\overrightarrow{add}_{a,d}[(\gamma_{1},\gamma
_{2},\overrightarrow{\delta})]\rightarrow(\gamma_{1},\gamma_{2}%
,\overrightarrow{\delta^{\prime}})$} & \multicolumn{1}{|l|}{$a\in\gamma_{1}$
and $d\in\delta_{00}\cup\delta_{01}$,}\\
\multicolumn{1}{|l|}{} & \multicolumn{1}{|l|}{$\delta_{00}^{\prime}%
=\delta_{00}-\{d\},\delta_{01}^{\prime}=\delta_{01}-\{d\},$}\\
\multicolumn{1}{|l|}{} & \multicolumn{1}{|l|}{$\delta_{10}^{\prime}=$
\texttt{if} $d\in\delta_{00}$ \texttt{then} $\delta_{10}\cup\{d\}$%
\ \texttt{else} $\delta_{10},$}\\
\multicolumn{1}{|l|}{} & \multicolumn{1}{|l|}{$\delta_{11}^{\prime}=$
\texttt{if} $d\in\delta_{01}$ \texttt{then} $\delta_{11}\cup\{d\}$%
\ \texttt{else} $\delta_{11}.$}\\\hline
\multicolumn{1}{|l|}{$\oplus\lbrack(\gamma_{1},\gamma_{2},\overrightarrow
{\delta}),(\gamma_{1}^{\prime},\gamma_{2}^{\prime},\overrightarrow
{\delta^{\prime}})]\rightarrow(\overline{\gamma}_{1},\overline{\gamma}%
_{2},\overrightarrow{\delta}\cup\overrightarrow{\delta^{\prime}})$} &
\multicolumn{1}{|l|}{$Disj(\overrightarrow{\delta}\cup\overrightarrow
{\delta^{\prime}})$, $\overline{\gamma}_{2}=\gamma_{2}\cup\gamma_{2}^{\prime
}\cup(\gamma_{1}\cap\gamma_{1}^{\prime})$,}\\
\multicolumn{1}{|l|}{} & \multicolumn{1}{|l|}{$\overline{\gamma}_{1}%
=(\gamma_{1}-(\gamma_{1}^{\prime}\cup\gamma_{2}^{\prime}))\cup(\gamma
_{1}^{\prime}-(\gamma_{1}\cup\gamma_{2})).$}\\\hline
\multicolumn{1}{|l|}{$\oplus\lbrack(\gamma_{1},\gamma_{2},\overrightarrow
{\delta}),(\gamma_{1}^{\prime},\gamma_{2}^{\prime},\overrightarrow
{\delta^{\prime}})]\rightarrow Error$} & \multicolumn{1}{|l|}{$\lnot
Disj(\overrightarrow{\delta}\cup\overrightarrow{\delta^{\prime}})$}\\\hline
\end{tabular}

\begin{center}
Table 1: Some transitions of $\mathcal{A}_{CT}$.
\end{center}

\emph{Remarks}: (1) We do not list the transitions relative
to$\ \overrightarrow{add}_{d,a}$\ because they are similar to those
of$\ \overrightarrow{add}_{a,d}$.

(2) For the transitions relative to$\ \overrightarrow{add}_{a,d}$, there are
three cases.\ It is clear that the conditions on $a$ and $d$ are mutually
exclusive and cover all possibilities.

(3) The transition $\overrightarrow{add}_{a,d}[(\gamma_{1},\gamma
_{2},\overrightarrow{\delta})]\rightarrow Error$\ is correct because the input
term is assumed irredundant. Without irredundancy, it may happen that
$a\in\gamma_{1}$ and $d\in\delta_{11}\cup\delta_{10}$ but $\overrightarrow
{add}_{a,d}$ has no effect because there are already edges from the $a$-vertex
to the $d$-vertex (or to several $d$-vertices), so that the $d$-vertex (or
vertices) still have indegree 1. In that case, the transition should yield
$(\gamma_{1},\gamma_{2},\overrightarrow{\delta}).$

(4) If we replace the transitions $\mathbf{a}\rightarrow(\{a\},\overrightarrow
{\emptyset})$\ and $\mathbf{d}\rightarrow(\emptyset,\emptyset
,\{d\},\overrightarrow{\emptyset})$ by $(\mathbf{a},w)\rightarrow
(\{a\},\overrightarrow{\emptyset})$\ and $(\mathbf{d},w^{\prime}%
)\rightarrow(\emptyset,\emptyset,\{d\},\overrightarrow{\emptyset})$
respectively for $w\in\{0,1\}^{p}$ \ and $w^{\prime}\in\{0,1\}^{m}$, we obtain
an automaton that checks the correctness of a term in $T(F_{C,D}^{(p,m)})$.

(5) Some states $(\gamma_{1},\gamma_{2},\overrightarrow{\delta})$ are not
accessible: for example, those such that $\gamma_{1}\cup\gamma_{2}=\emptyset$
and $\delta_{01}\cup\delta_{10}\cup\delta_{11}\neq\emptyset.$

\bigskip

If $C\cup D$ is finite, $k=\left\vert C\right\vert $ and $\ell=\left\vert
D\right\vert $, the number of states is $3^{k}.5^{\ell}+1$ and the size of a
state is $O(k+\ell)$. Each transition is computable in time $O(k+\ell)$ (our
time complexity evaluations are based on straightforward data structures). We
obtain a linear FPT-FA.

\bigskip

As noted previously, $h^{-1}(\mathcal{A}_{CT})=\mathcal{A}_{Subgraph(X,U)}$
where\ $h$ maps $(\mathbf{a},1)$ to $\mathbf{a}$ and $(\mathbf{a},0)$ to
$\varnothing$ for each $a\in C\cup D$.

\subsection{Adjacency in incidence graphs}

In the case of MSO\ graph properties reviewed in Section 2 the atomic formula
$edg(X,Y)$ is checked by an FA over $F_{C}^{(2)}$\ that has\ $k^{2}+k+3$
states if $C$ is finite of cardinality $k$.\ This construction is easily
applicable to the atomic formula $inc(X,U)$ (resp.\ $inc(U,X)$) stating that
$X$ consists of one vertex $x$ and $U$ of one edge $u$ whose tail
(resp.\ head) is $x$. We will describe $\mathcal{A}_{inc(X,U)}$\ at the end of
the section for purpose of comparison.\ 

\bigskip

In MSO\ formulas expressing properties of $Inc(G)$, the property $edg(X,Y)$ is
no longer atomic; it is expressed by $\exists U.(inc(X,U)\wedge inc(U,Y)). $
We can apply the general construction of \cite{BCID12, CouEng} to this
formula, but, in view of practical constructions, it is useful to define
directly an automaton $\mathcal{A}_{edg(X,Y)}$ over $F_{C,D}^{(2,0)}.$

\bigskip

\emph{The FA }$\mathcal{A}_{edg(X,Y)\ }$

We now construct an FA $\mathcal{A}_{edg(X,Y)}$ intended to run on correct and
irredundant terms in $T(F_{C,D}^{(2,0)})$.\ As for irredundancy, we will
assume that correctness is guaranteed by the parsing algorithm.

The states of $\mathcal{A}_{edg(X,Y)}$ are $Ok$, $Error$ and the tuples
$(\gamma_{1},\gamma_{2},\overrightarrow{\delta})\in\mathcal{P}_{\leq1}%
(C)^{2}\times\mathcal{P}_{f}(D)^{3}$ such that the components of
$\overrightarrow{\delta}=(\delta,\delta_{1},\delta_{2})$ verify the condition
$\delta_{1}\cup\delta_{2}\subseteq\delta$.

Let $t\in T(F_{C,D}^{(2,0)})$ be a correct and irredundant term.\ It defines
an incidence graph $val(t)=Inc(G)$ and two sets of vertices $X,Y$\ of $G$.
Every subterm $t^{\prime}$ of $t$ is irredundant and defines a bipartite graph
$val(t^{\prime})$ that we consider as a subgraph of $val(t)$ (by our
\emph{Convention about subterms}, cf.\ Section 2).\ It may not be an incidence
graph, because some $D$-vertices may be of indegree or outdegree 0. Let
$X^{\prime}=X\cap V_{val(t^{\prime})},Y^{\prime}=Y\cap V_{val(t^{\prime})}%
.$Their sets of labels are $\pi_{val(t^{\prime})}(X^{\prime})$ and
$\pi_{val(t^{\prime})}(Y^{\prime})$, both subsets of $C$. To simplify notation
we will denote $\pi_{val(t^{\prime})}$\ by $\pi$ and $\rightarrow
_{val(t^{\prime})}$ (the edge relation) by $\rightarrow$. At the root of
$t^{\prime},$ $\mathcal{A}_{edg(X,Y)}$ reaches the following state:

\begin{quote}
$Error$ if and only if $X^{\prime}$ or $Y^{\prime}$ has cardinality 2 or more,

$Ok$ if and only if $X^{\prime}=\{x\}$, $Y^{\prime}=\{y\}$ for some $x,y$ such
that $x\rightarrow u\rightarrow y$ for some $D$-vertex $u$\ (so that
$x\rightarrow_{G}y$),

$(\gamma_{1},\gamma_{2},\overrightarrow{\delta})$ otherwise, and we have :

$\gamma_{1}=\pi(X^{\prime})$ and $\left\vert X^{\prime}\right\vert \leq1,$

$\gamma_{2}=\pi(Y^{\prime})$ and $\left\vert Y^{\prime}\right\vert \leq1,$

$\delta$ is the set of labels of $D$-vertices,

$\delta_{1}=\pi(Out(X^{\prime}))\subseteq\delta,$

$\delta_{2}=\pi(In(Y^{\prime}))\subseteq\delta,$

where $Out(X^{\prime})$ (resp.\ $In(Y^{\prime}))$\ is the set of $D$-vertices
$u$ such that $X^{\prime}\rightarrow u$ (resp. $u\rightarrow Y^{\prime}).$
\end{quote}

The accepting state is $Ok$.\ %

\begin{tabular}
[c]{|c|c|}\hline
Transitions & Conditions\\\hline
\multicolumn{1}{|l|}{$\varnothing\rightarrow\overrightarrow{\emptyset}$} &
\multicolumn{1}{|l|}{}\\\hline
\multicolumn{1}{|l|}{$(\mathbf{a},00)\rightarrow\overrightarrow{\emptyset}\ $}
& \\\hline
\multicolumn{1}{|l|}{$(\mathbf{a},10)\rightarrow(\{a\},\overrightarrow
{\emptyset})$} & \multicolumn{1}{|l|}{}\\\hline
\multicolumn{1}{|l|}{$(\mathbf{a},01)\rightarrow(\emptyset
,\{a\},\overrightarrow{\emptyset})$} & \multicolumn{1}{|l|}{}\\\hline
\multicolumn{1}{|l|}{$(\mathbf{a},11)\rightarrow(\{a\},\{a\},\overrightarrow
{\emptyset})$} & \multicolumn{1}{|l|}{}\\\hline
\multicolumn{1}{|l|}{$\mathbf{d}\rightarrow(\emptyset,\emptyset
,\{d\},\overrightarrow{\emptyset})$} & \multicolumn{1}{|l|}{}\\\hline
\multicolumn{1}{|l|}{$\overrightarrow{add}_{a,d}[(\gamma_{1},\gamma
_{2},\overrightarrow{\delta})]\rightarrow(\gamma_{1},\gamma_{2}%
,\overrightarrow{\delta})$} & \multicolumn{1}{|l|}{$a\notin\gamma_{1}$ or
$d\notin\delta$}\\\hline
\multicolumn{1}{|l|}{$\overrightarrow{add}_{a,d}[(\gamma_{1},\gamma
_{2},\overrightarrow{\delta})]\rightarrow Ok$} & \multicolumn{1}{|l|}{$a\in
\gamma_{1}$ and $d\in\delta_{2}$}\\\hline
\multicolumn{1}{|l|}{$\overrightarrow{add}_{a,d}[(\gamma_{1},\gamma
_{2},\overrightarrow{\delta})]\rightarrow(\gamma_{1},\gamma_{2},\delta
,\delta_{1}^{\prime},\delta_{2})$} & \multicolumn{1}{|l|}{$a\in\gamma_{1}$,
$d\in\delta-\delta_{2}$, $\delta_{1}^{\prime}:=\delta_{1}\cup\{d\}$}\\\hline
\multicolumn{1}{|l|}{$\overrightarrow{add}_{a,d}[Ok]\rightarrow Ok$} &
\\\hline
\multicolumn{1}{|l|}{$relab_{a\rightarrow b}[(\gamma_{1},\gamma_{2}%
,\overrightarrow{\delta})]\rightarrow(\gamma_{1}^{\prime},\gamma_{2}^{\prime
},\overrightarrow{\delta})$} & \multicolumn{1}{|l|}{$\gamma_{1}^{\prime
}:=\gamma_{1}[a\rightarrow b],\gamma_{2}^{\prime}:=\gamma_{2}[a\rightarrow
b]$}\\\hline
\multicolumn{1}{|l|}{$relab_{a\rightarrow b}[Ok]\rightarrow Ok$} & \\\hline
\multicolumn{1}{|l|}{$relab_{d\rightarrow e}[(\gamma_{1},\gamma_{2}%
,\overrightarrow{\delta})]\rightarrow(\gamma_{1},\gamma_{2},\overrightarrow
{\delta^{\prime}})$} & \multicolumn{1}{|l|}{$\overrightarrow{\delta^{\prime}%
}:=\overrightarrow{\delta}[d\rightarrow e]$}\\\hline
\multicolumn{1}{|l|}{$\oplus\lbrack Ok,Ok]\rightarrow Error$} &
\multicolumn{1}{|l|}{}\\\hline
\multicolumn{1}{|l|}{$\oplus\lbrack(\gamma_{1},\gamma_{2},\overrightarrow
{\delta}),Ok]\rightarrow Error$} & \multicolumn{1}{|l|}{$\gamma_{1}%
\neq\emptyset$ or $\gamma_{2}\neq\emptyset$}\\\hline
\multicolumn{1}{|l|}{$\oplus\lbrack(\gamma_{1},\gamma_{2},\overrightarrow
{\delta}),Ok]\rightarrow Ok$} & \multicolumn{1}{|l|}{$\gamma_{1}=\gamma
_{2}=\emptyset$}\\\hline
\multicolumn{1}{|l|}{$\oplus\lbrack(\gamma_{1},\gamma_{2},\overrightarrow
{\delta}),(\gamma_{1}^{\prime},\gamma_{2}^{\prime},\overrightarrow
{\delta^{\prime}})]\rightarrow Error$} & \multicolumn{1}{|l|}{$\left\vert
\gamma_{1}\right\vert +\left\vert \gamma_{1}^{\prime}\right\vert \geq2$ or
$\left\vert \gamma_{2}\right\vert +\left\vert \gamma_{2}^{\prime}\right\vert
\geq2$}\\\hline
\multicolumn{1}{|l|}{$\oplus\lbrack(\gamma_{1},\gamma_{2},\overrightarrow
{\delta}),(\gamma_{1}^{\prime},\gamma_{2}^{\prime},\overrightarrow
{\delta^{\prime}})]$} & \multicolumn{1}{|l|}{otherwise}\\
\multicolumn{1}{|l|}{$\ \ \ \ \ \ \ \ \ \ \ \ \ \ \ \ \ \ \rightarrow
(\gamma_{1}\cup\gamma_{1}^{\prime},\gamma_{2}\cup\gamma_{2}^{\prime
},\overrightarrow{\delta}\cup\overrightarrow{\delta})$} &
\multicolumn{1}{|l|}{}\\\hline
\end{tabular}

\begin{center}
Table 2\ : Some transitions of $\mathcal{A}_{edg(X,Y)}$
\end{center}

\bigskip

\emph{Remarks} : (1) The transitions $\oplus\lbrack(\emptyset,\emptyset
,\overrightarrow{\delta}),Ok]\rightarrow Ok$ "loose" the value
$\overrightarrow{\delta}.$ This value is not needed after $Ok$ is
obtained.\ The only thing that remains to be checked is that the sets $X,Y$
are singletons. It follows that we have transitions such as $\oplus\lbrack
Ok,Ok]\rightarrow Error$ \ and $\oplus\lbrack(\gamma_{1},\gamma_{2}%
,\overrightarrow{\delta}),Ok]\rightarrow Error$ if $\gamma_{1}\neq\emptyset$
or $\gamma_{2}\neq\emptyset.$ The state $Ok$ is not a sink.

(2) Let us comment on $\delta_{2}$.\ A $D$-vertex $u$ such that $u\rightarrow
Y^{\prime}$ corresponds either to a partially defined edge with head in
$Y^{\prime}$ and whose tail is not yet found or to an edge from a vertex not
in $X^{\prime}$ to a vertex in $Y^{\prime}$.\ The set $\delta_{2}$ is the set
of their labels. The transition $\overrightarrow{add}_{a,d}[(\gamma_{1}%
,\gamma_{2},\overrightarrow{\delta})]\rightarrow Ok$ when $\gamma_{1}=\{a\}$
and $d\in\delta_{2}$\ is correct because, since $t $ is assumed to be a
correct term, a $D$-vertex $u$ with label $d$ cannot correspond to an edge
from a vertex not in $X^{\prime}$ to a vertex in $Y^{\prime}$.\ Hence,
$u\rightarrow y\in Y^{\prime}$ and the operation $\overrightarrow{add}_{a,d}$
creates an edge from the vertex of $X^{\prime}$ to $y$.

(3) Some states \ $(\gamma_{1},\gamma_{2},\overrightarrow{\delta})$ are not
accessible, for example, those such that $\gamma_{1}=\emptyset$ and
$\delta_{1}\neq\emptyset.$

\bigskip

If $C\cup D$ is finite, $k=\left\vert C\right\vert $ and $\ell=\left\vert
D\right\vert $, the number of states is $(k+1)^{2}.5^{\ell}+2$ and the size of
a state is $O(\log(k)+\ell))$. Each transition is computable in time
$O(\log(k)+\ell))$. We obtain a linear FPT-FA.

\bigskip

\emph{Comparison with the automaton constructed with Theorem 4.}

\bigskip

The automaton $\mathcal{A}_{inc(X,U)}$ over $F_{C,D}^{(1,1)}$ obtained by a
straightforward adaptation of the automaton $\mathcal{A}_{edg(X,Y)}$ over
$F_{C}^{(2)}$ defined in \cite{BCID12, CouEng}.\ It has states $Ok$, $Error$
and the pairs $(\gamma,\delta)$ in $\mathcal{P}_{\leq1}(C)\times
\mathcal{P}_{\leq1}(D).\ $

Let $t\ast(X,U)\in T(F_{C,D}^{(1,1)})$ be correct and irredundant.\ It defines
an incidence graph $val(t)=Inc(G)$, a set $X$\ of $C$-vertices and a set $U$
of $D$-vertices.\ Every subterm $t^{\prime}$ of $t$ defines a bipartite graph
$val(t^{\prime}).$ Let $X^{\prime}=X\cap V_{val(t^{\prime})},U^{\prime}=U\cap
V_{val(t^{\prime})}.$ At the root of $t^{\prime}\ast(X^{\prime},U^{\prime}),$
$\mathcal{A}_{inc(X,U)}$ reaches the following state:

\begin{quote}
$Error$ if and only if $X^{\prime}$ or $U^{\prime}$ has cardinality 2 or more,

$Ok$ if and only if $X^{\prime}=\{x\}$, $U^{\prime}=\{u\}$ and $x\rightarrow
_{val(t^{\prime})}u$,

$(\gamma,\delta)$ otherwise where

$\gamma=\pi_{val(t^{\prime})}(X^{\prime})$ and $\left\vert X^{\prime
}\right\vert \leq1,$

$\delta=\pi_{val(t^{\prime})}(U^{\prime})$ and $\left\vert U^{\prime
}\right\vert \leq1.$
\end{quote}

The accepting state is $Ok$ and the transitions are easy to define.\ If $C\cup
D$ is finite, $k=\left\vert C\right\vert $ and $\ell=\left\vert D\right\vert
$, the number of states is $(k+1).(\ell+1)+2.$

Similarly, we have $\mathcal{A}_{inc(U,Y)}$ with same set of states, except
that we write $(\delta,\gamma)=(\pi_{val(t^{\prime})}(U^{\prime}%
),\pi_{val(t^{\prime})}(Y^{\prime}))\in\mathcal{P}_{\leq1}(D)\times
\mathcal{P}_{\leq1}(C)$ instead of $(\gamma,\delta)$.

Let us now consider $edg(X,Y)$ defined by $\exists U.(inc_{1}(X,U,Y)\wedge
inc_{2}(X,U,Y))$ where $inc_{1}(X,U,Y)$ is defined as $inc(X,U)$ with extra
(useless) variable $Y$\ and $inc_{2}(X,U,Y))$ where $inc_{2}(X,U,Y)$ is
defined as $inc(U,Y)$.\ These useless variables are added so that the two
parts of the conjunction $inc_{1}(X,U,Y)\wedge inc_{2}(X,U,Y)$ have the same
free variables.\ 

The automaton $\mathcal{B}:=\mathcal{A}_{inc_{1}(X,U,Y)\wedge inc_{2}(X,U,Y)}$
over $F_{C,D}^{(2,1)}$ is thus the product of $\mathcal{A}_{inc_{1}(X,U,Y)}$
and $\mathcal{A}_{inc_{2}(X,U,Y)}$.\ Furthermore, $\mathcal{A}_{inc_{1}%
(X,U,Y)}$ and $\mathcal{A}_{inc_{2}(X,U,Y)}$ have the sames states as,
respectively, $\mathcal{A}_{inc(X,U)}$ and $\mathcal{A}_{inc(U,Y)}$. (See the
remarks on variable substitutions after Theorem 5). The construction of
Theorem 5 replaces $\mathcal{B}$\ by a nondeterministic FA\ $\mathcal{C}%
$\ over $F_{C,D}^{(2,0)}$ that characterizes $\exists U.(inc_{1}(X,U,Y)\wedge
inc_{2}(X,U,Y))$. The FA $\mathcal{B}$ has $((k+1).(\ell+1)+2)^{2}$ states if
$k=\left\vert C\right\vert $ and $\ell=\left\vert D\right\vert $, which
gives\ more than $2^{k^{2}.\ell^{2}}$ states for the deterministic automaton
$\det(\mathcal{C})$ obtained from $\mathcal{C}$.\ However, some states of
$\mathcal{B}$ are inaccessible, for instance, those of the form $((\gamma
,\delta),(\delta^{\prime},\gamma^{\prime}))$ where $\delta\neq\delta^{\prime
}.$\ Furthermore, the states of $\det(\mathcal{C})$ are more complicated to
write than those of $\mathcal{A}_{edg(X,Y)}$ are in the set $\{Ok,Error\}\cup
(\mathcal{P}_{\leq1}(C)^{2}\times\mathcal{P}_{f}(D)^{3})$ whereas those of
$\det(\mathcal{C})$ are in $\mathcal{P}_{f}([\{Ok,Error\}\cup(\mathcal{P}%
_{\leq1}(C)\times\mathcal{P}_{\leq1}(D))]^{2}).$

This observation motivates our interest for "direct constructions" of FA for
properties related to adjacency, as these properties are defined with
$edg(X,Y)$.

\bigskip

\subsection{Links and domination}

\bigskip

We consider the following four properties based on adjacency, ordered by
increasing complexity, measured by the sizes of the automata we will construct
($G$ is the graph whose incidence graph is $val(t)$ and $t$ the given correct
term) :

\begin{quote}
$Link^{\exists\exists}(X,Y)$ meaning that $X\rightarrow_{G}Y,$ i.e.,
$x\rightarrow_{G}y$ for some $x\in X$ and $y\in Y$,

$Link^{\forall\exists}(X,Y)$ meaning that for all $x\in X$ there is $y\in Y $
such that $x\rightarrow_{G}y$, ($Y$ \emph{dominates} $X$\ for $\leftarrow_{G}$),

$Link^{\forall\forall}(X,Y)$ meaning that $x\rightarrow_{G}y$ for all\ $x\in
X$ and $y\in Y$,

$Link^{\exists\forall}(X,Y)$ meaning that there is $x\in X$ such that
$x\rightarrow_{G}y$ for all $y\in Y$ ($x$ \emph{dominates} $Y$\ for
$\rightarrow_{G}$).
\end{quote}

\bigskip

The property that $X$\ induces a complete directed graph (with loops on all
vertices) is expressed by $Link^{\forall\forall}(X,X)$. That $X$\ is stable,
i.e., that the induced graph $G[X]$ has no edge, is expressed by $\lnot
Link^{\exists\exists}(X,X)$.

The logical expressions of these properties by MSO\ formulas interpreted in
$\left\langle V_{G}\cup E_{G},inc_{G}\right\rangle $ have the following
respective quantifier structures: $\exists\exists\exists,\forall\exists
\exists,$ $\ \forall\forall\exists$ \ and\ $\exists\forall\exists$ with 0, 1,1
and 2 quantifier alternations.\ We will construct FA or sketch their
constructions.\ Their sets of states will reflect the differences of
quantifier alternations. Notation is as for the definition of $\mathcal{A}%
_{edg(X,Y)}$. All FA\ will be intended to run on correct and irredundant terms
in $T(F_{C,D}^{(2,0)})$.

\bigskip

\emph{The FA }$\mathcal{A}_{Link^{\exists\exists}(X,Y)}$

It is similar to $\mathcal{A}_{edg(X,Y)}$ with some interesting
differences.\ It is simpler to describe because it need not check that $X$ and
$Y$ are singletons, however, it has more states.\ Its states are the accepting
sink $Success$ and the tuples $(\gamma_{1},\gamma_{2},\overrightarrow{\delta
})\in\mathcal{P}_{f}(C)^{2}\times\mathcal{P}_{f}(D)^{3}$ such that $\delta
_{1}\cup\delta_{2}\subseteq\delta$. As above for $\mathcal{A}_{edg(X,Y)}$, if
$t^{\prime}\ast(X^{\prime},Y^{\prime})$\ is a subterm of a correct irredundant
term $t\ast(X,Y)\in T(F_{C,D}^{(2,0)})$ where $X^{\prime}=X\cap
V_{val(t^{\prime})}$ and $Y^{\prime}=Y\cap V_{val(t^{\prime})},$\ then
$\mathcal{A}_{Link^{\exists\exists}(X,Y)}$ reaches the following state at the
root of $t^{\prime}\ast(X^{\prime},Y^{\prime})$\ :

\begin{quote}
$Success$ if and only if $x\rightarrow u\rightarrow y$ for some $x\in
X^{\prime}$ and $y\in Y^{\prime}$ and $u$ (so that $X^{\prime}\rightarrow
_{G}Y^{\prime}$); otherwise

$(\gamma_{1},\gamma_{2},\overrightarrow{\delta})$ where :

$\qquad\gamma_{1}=\pi(X^{\prime})\subseteq C,$

$\qquad\gamma_{2}=\pi(Y^{\prime})\subseteq C,$

$\qquad\delta$ is the set of labels of $D$-vertices,

$\qquad\delta_{1}=\pi(Out(X^{\prime}))\subseteq\delta,$

$\qquad\delta_{2}=\pi(In(Y^{\prime}))\subseteq\delta.$

\bigskip
\end{quote}

Some transitions are listed in Table 3 (the others are easy to define or
similar to those for $\mathcal{A}_{edg(X,Y)}$).%

\begin{tabular}
[c]{|c|c|}\hline
Transitions & Conditions\\\hline\hline
\multicolumn{1}{|l|}{$\overrightarrow{add}_{a,d}[(\gamma_{1},\gamma
_{2},\overrightarrow{\delta})]\rightarrow(\gamma_{1},\gamma_{2}%
,\overrightarrow{\delta})$} & \multicolumn{1}{|l|}{$a\notin\gamma_{1}$ or
$d\notin\delta$}\\\hline
\multicolumn{1}{|l|}{$\overrightarrow{add}_{a,d}[(\gamma_{1},\gamma
_{2},\overrightarrow{\delta})]\rightarrow Success$} &
\multicolumn{1}{|l|}{$a\in\gamma_{1}$ and $d\in\delta_{2}$}\\\hline
\multicolumn{1}{|l|}{$\overrightarrow{add}_{a,d}[(\gamma_{1},\gamma
_{2},\overrightarrow{\delta})]\rightarrow(\gamma_{1},\gamma_{2},\delta
,\delta_{1}\cup\{d\},\delta_{2})$} & \multicolumn{1}{|l|}{$a\in\gamma_{1}$,
$d\in\delta-\delta_{2}$}\\\hline
\multicolumn{1}{|l|}{$\oplus\lbrack(\gamma_{1},\gamma_{2},\overrightarrow
{\delta}),(\gamma_{1}^{\prime},\gamma_{2}^{\prime},\overrightarrow
{\delta^{\prime}})]$} & \multicolumn{1}{|l|}{}\\
\multicolumn{1}{|l|}{$\ \ \ \ \ \ \ \ \ \ \ \ \ \ \ \ \ \ \ \ \ \ \ \rightarrow
(\gamma_{1}\cup\gamma_{1}^{\prime},\gamma_{2}\cup\gamma_{2}^{\prime
},\overrightarrow{\delta}\cup\overrightarrow{\delta^{\prime}})$} &
\multicolumn{1}{|l|}{}\\\hline
\end{tabular}

\begin{center}
Table 3: Some transitions of $\mathcal{A}_{Link^{\exists\exists}(X,Y)}$.
\end{center}

The accepting state is the sink $Success$. All transitions with $Success$ on
the left yield $Success$. (In $\mathcal{A}_{edg(X,Y)}$, we use a state $Ok$,
that looks like $Success$ but is not a sink.) See Remark (2) relative to
$\mathcal{A}_{edg(X,Y)}$ to verify the validity of the transition to $Success$
in this table. There is no $Error$ state. \ 

\bigskip

If $C\cup D$ is finite, $k=\left\vert C\right\vert $ and $\ell=\left\vert
D\right\vert $, the number of states is $4^{k}.5^{\ell}+1.$ These states have
size $O(k+\ell)$. Each transition is computed in time $O(k+\ell)$. We obtain a
linear FPT-FA.

\bigskip

\emph{The FA }$\mathcal{A}_{Link^{\forall\forall}(X,Y)}$

\bigskip

Notation is as for the previous automaton. The state at the root of
$t^{\prime}$ will contain the relation :

\begin{quote}
$\theta:=\{(\pi(x),\pi(y))\mid x\in X^{\prime},y\in Y^{\prime}$ \ and
$\nexists u.(x\rightarrow u\rightarrow y)\},$
\end{quote}

and will be accepting if and only if this relation is empty, because $\theta$
indicates the existence of "missing" edges. \ However, additional information
will be needed for the construction of transition rules.

A state of $\mathcal{A}_{Link^{\forall\forall}(X,Y)}$ is a 4-tuple of finite
sets $(\delta,\Delta,\Lambda,\Theta)$ such that:

\begin{quote}
$\delta\subseteq D,$

$\Delta\subseteq C\times\mathcal{P}_{f}(\delta)$, $\Lambda\subseteq
\mathcal{P}_{f}(\delta)\times C$ and

$\Theta\subseteq\Delta\times\Lambda\subseteq C\times\mathcal{P}_{f}%
(D)\times\mathcal{P}_{f}(D)\times C.$
\end{quote}

The state at the root of $t^{\prime}\ast(X^{\prime},Y^{\prime})$ is
$(\delta,\Delta,\Lambda,\Theta)$\ such that :

\begin{quote}
$\delta$ is the set of labels of $D$-vertices,

$\Delta=\{(\pi(x),\pi(Out(x))\mid x\in X^{\prime}\},$

$\Lambda=\{(\pi(In(y)),\pi(y))\mid y\in Y^{\prime}\},$

$\Theta=\{(\pi(x),\pi(Out(x)),\pi(In(y)),\pi(y))\mid x\in X^{\prime},$

$\qquad\qquad\qquad y\in Y^{\prime}$ and $\nexists u.(x\rightarrow
u\rightarrow y)\}.$
\end{quote}

\bigskip

A tuple $(a,\eta,\eta^{\prime},b)$ in $\Theta$ encodes the following
information about a "missing edge" from some $x\in X^{\prime}$ to some $y\in
Y^{\prime}$: $a=\pi(x)$, $b=\pi(y),$ $\eta$ contains the labels of the
partially defined edges with tail $x$ and similarly, $\eta^{\prime}$ contains
the labels of those, partially defined, with head $y$. An edge from $x$ to $y$
in the graph $G$ whose incidence graph is $val(t)$ can be created by
$\overrightarrow{add}_{a,d}$ if $d\in\eta^{\prime}$ or by $\overrightarrow
{add}_{d,b}$ if $d\in\eta$ (assuming that labels $a,d$ are not modified above
$t^{\prime})$.

\bigskip

\emph{Remark }: The states of $\mathcal{A}_{Link^{\forall\forall}(X,Y)}$
contain more information than those of $\mathcal{A}_{Link^{\exists\exists
}(X,Y)}$ because $\pi(X^{\prime})=\{a\in C\mid(a,\eta)\in\Delta$ for some
$\eta\},$ $\pi(Out(X^{\prime}))$\ is the union of the sets $\eta$ such that
$(a,\eta)\in\Delta$ for some $a$ and similarly for $\pi(Y^{\prime})$ and
$\pi(In(Y^{\prime})).$\ Hence, the state of $\mathcal{A}_{Link^{\exists
\exists}(X,Y)}$ at some position $u$ can be computed from that of
$\mathcal{A}_{Link^{\forall\forall}(X,Y)}$\ at $u$.\ One could formalize that
by the existence of a homomorphism : $\mathcal{A}_{Link^{\forall\forall}%
(X,Y)}\rightarrow\mathcal{A}_{Link^{\exists\exists}(X,Y)}$.\ $\square$

\bigskip

The accepting states are those such that $\Theta=\emptyset$ because, at the
root of a correct term $t$, $\theta$ is the set of pairs $(a,b)$ such that
$(a,\eta,\eta^{\prime},b)\in\Theta$ for some $\eta,\eta^{\prime}.$ Transitions
are as follows.\ We begin with the easier cases.

\begin{quote}
$\varnothing\rightarrow\overrightarrow{\emptyset}$

$(\mathbf{a},11)\rightarrow(\emptyset,\{(a,\emptyset)\},\{(\emptyset
,a)\},\{(a,\emptyset,\emptyset,a)\}),$

$(\mathbf{a},10)\rightarrow(\emptyset,\{(a,\emptyset)\},\overrightarrow
{\emptyset}),$

$(\mathbf{a},01)\rightarrow(\emptyset,\{(\emptyset,a)\},\overrightarrow
{\emptyset}),$

$(\mathbf{a},00)\rightarrow\overrightarrow{\emptyset},$

$\mathbf{d}\rightarrow(\{d\},\overrightarrow{\emptyset}).$
\end{quote}

\bigskip

Transitions for relabellings are straightforward:

$relab_{x\rightarrow y}[(\delta,\Delta,\Lambda,\Theta)]\rightarrow
(\delta^{\prime},\Delta^{\prime},\Lambda^{\prime},\Theta^{\prime})$ where
$\delta^{\prime}:=\delta\lbrack x\rightarrow y]$ \ and similarly for the other components.

\bigskip

Transitions for disjoint union are as follows:

\begin{quote}
$\oplus\lbrack(\delta_{1},\Delta_{1},\Lambda_{1},\Theta_{1}),(\delta
_{2},\Delta_{2},\Lambda_{2},\Theta_{2})]\rightarrow$

$\qquad(\delta_{1}\cup\delta_{2},\Delta_{1}\cup\Delta_{2},\Lambda_{1}%
\cup\Lambda_{2},\Theta_{1}\cup\Theta_{2}\cup\Theta^{\prime})$\ where

$\qquad\Theta^{\prime}:=\{(a,\eta,\eta^{\prime},b)\mid((a,\eta)\in\Delta_{1}$
and $(\eta^{\prime},b)\in\Lambda_{2})$

\qquad\qquad\qquad\qquad or $((a,\eta)\in\Delta_{2}$ and $(\eta^{\prime}%
,b)\in\Lambda_{1}))\}.$

\bigskip
\end{quote}

Transitions for edge addition are as follows:

$\overrightarrow{add}_{a,d}[(\delta,\Delta,\Lambda,\Theta)]\rightarrow q$
\ where we have:

\begin{quote}
$q:=(\delta,\Delta,\Lambda,\Theta)$ if $a$ does not occur in $\Delta$ or
$d\notin\delta,$

$q:=(\delta,\Delta^{\prime},\Lambda,\Theta^{\prime})$ otherwise, where $:$

$\qquad\Delta^{\prime}$\ is defined from $\Delta$\ by replacing each pair
$(a,\eta)$ by $(a,\eta\cup\{d\})$;

$\qquad\Theta^{\prime}$ \ is defined from in $\Theta$ as follows :

\qquad\qquad a tuple of the form $(a,\eta,\eta^{\prime},b)$, for any $b\in C$
is deleted

\qquad\qquad if $d\in\eta^{\prime}$; otherwise it is replaced by $(a,\eta
\cup\{d\},\eta^{\prime},b).$
\end{quote}

\bigskip

In the transitions for\emph{\ }$\overrightarrow{add}_{a,d}$, we cannot have
$a$\ and $d$ occurring both in $\Delta$ because the input term $t$ is assumed
irredundant and correct. If $d$ belongs to $\eta^{\prime}$, then it labels
vertices of indegree 0 because $t$ is correct and irredundant.

\bigskip

If $C\cup D$ is finite, $k=\left\vert C\right\vert $ and $\ell=\left\vert
D\right\vert $, the number of states is at most $2^{\ell}.5^{k^{2}4^{\ell}}$.
However, there are less accessible states.\ Counting them precisely seems
difficult and is actually not important for our use of FA. The size of a state
is $O(k^{2}.4^{\ell})$. We obtain a linear FPT-FA.

\bigskip

\emph{The FA\ }$\mathcal{A}_{Link^{\forall\exists}(X,Y)}$

\bigskip

The construction is similar to the previous one.\ The set of states is a bit
smaller, although of similar type.\ Notation is as in the previous cases. The
states are sets of 5-tuples of sets $(\gamma,\delta,\lambda,\Delta,\Theta
)\in\mathcal{P}_{f}(C)\times\mathcal{P}_{f}(D)^{2}\times\mathcal{P}%
_{f}(C\times\mathcal{P}_{f}(D))^{2}$ such that:

\begin{quote}
$\gamma\subseteq C,\delta\subseteq D,\lambda\subseteq\delta$ and
$\Theta\subseteq\Delta\subseteq C\times\mathcal{P}_{f}(\delta)$.
\end{quote}

The state at the root of $t^{\prime}\ast(X^{\prime},Y^{\prime})$ \ is
$(\gamma,\delta,\lambda,\Delta,\Theta)$\ such that :

\begin{quote}
$\gamma=\pi(Y^{\prime}),$

$\delta$ is the set of labels of $D$-vertices,

$\lambda=\pi(In(Y^{\prime}))\subseteq\delta,$

$\Delta=\{(\pi(x),\pi(Out(X^{\prime})))\mid x\in X^{\prime}\},$

$\Theta=\{(\pi(x),\pi(Out(X^{\prime})))\mid x\in X^{\prime}\ $and $\nexists
u.(x\rightarrow u\rightarrow Y^{\prime})\}.$
\end{quote}

\bigskip

The accepting states are those such that $\Theta=\emptyset$.

\bigskip

Transitions for disjoint union are as follows:

\begin{quote}
$\oplus\lbrack(\gamma_{1},\delta_{1},\lambda_{1},\Delta_{1},\Theta
_{1}),(\gamma_{2},\delta_{2},\lambda_{2},\Delta_{2},\Theta_{2})]\rightarrow$

$\qquad\qquad\qquad(\gamma_{1}\cup\gamma_{2},\delta_{1}\cup\delta_{2}%
,\lambda_{1}\cup\lambda_{2},\Delta_{1}\cup\Delta_{2},\Theta_{1}^{\prime}%
\cup\Theta_{2}^{\prime})$.
\end{quote}

where:

\begin{quote}
$\Theta_{1}^{\prime}:=$ \texttt{if} $\gamma_{2}=\emptyset$ \ \texttt{then
}$\Theta_{1}$\ \texttt{else }$\Delta_{1}$, and

$\Theta_{2}^{\prime}:=$ \texttt{if} $\gamma_{1}=\emptyset$\ \texttt{then
}$\Theta_{2}$\ \texttt{else}  $\Delta_{2}$.
\end{quote}

\bigskip

The most complicated transitions are for edge addition:

$\overrightarrow{add}_{a,d}[(\gamma,\delta,\lambda,\Delta,\Theta)]\rightarrow
q$ \ where the following holds:

\begin{quote}
if $a$ \ does not occur in $\Delta$ or $d\notin\delta,$ then $q:=(\gamma
,\delta,\lambda,\Delta,\Theta),$

otherwise $q:=(\gamma,\delta,\lambda,\Delta^{\prime},\Theta^{\prime}), $ where
$\Delta^{\prime},\Theta^{\prime}$ are defined as follows from $\Delta$ and
$\Theta:$

\qquad a pair $(a,\eta)$ in $\Delta$ is replaced by $(a,\eta\cup\{d\})$,

\qquad a pair $(a,\eta)$ in $\Theta$ is deleted if $d\in\lambda$ and replaced
by $(a,\eta\cup\{d\})$ otherwise.
\end{quote}

\bigskip

$\overrightarrow{add}_{d,b}[(\gamma,\delta,\lambda,\Delta,\Theta)]\rightarrow
q$ \ where the following holds:

\begin{quote}
if $b\notin\gamma$ or $d\notin\delta,$ then $q=(\gamma,\delta,\lambda
,\Delta,\Theta),$

otherwise $q:=(\gamma,\delta,\lambda\cup\{d\},\Delta,\Theta^{\prime})$ where
$\ \Theta^{\prime}$ is defined from $\Theta$ by deleting each pair $(a,\eta)$
such that if $d\in\eta$.\ 
\end{quote}

\bigskip

If $C\cup D$ is finite, $k=\left\vert C\right\vert $ and $\ell=\left\vert
D\right\vert $, the number of states is bounded by $2^{k}3^{\ell}%
.3^{k.2^{\ell}}.$ The size of a state is $O(k.2^{\ell})$ and the computation
time of a transition is $O(k.2^{\ell})$. We obtain a linear FPT-FA.

\bigskip

\emph{The FA }$\mathcal{A}_{Link^{\exists\forall}(X,Y)}$

\bigskip

This automaton is more complicated than the two previous ones, as it checks a
formula with two quantifier alternations instead of one. Its states are
triples of finite sets $(\delta,\Lambda,\Xi)$ such that:

\begin{quote}
$\delta\subseteq D,$

$\Lambda\subseteq\mathcal{P}_{f}(\delta)\times C,$

$\Xi\subseteq C\times\mathcal{P}_{f}(\delta)\times\mathcal{P}_{f}%
(\Lambda)\subseteq C\times\mathcal{P}_{f}(D)\times\mathcal{P}_{f}%
(\mathcal{P}_{f}(D)\times C).$
\end{quote}

The state at the root of $t^{\prime}\ast(X^{\prime},Y^{\prime})$ is
$(\delta,\Lambda,\Xi)$\ such that :

\begin{quote}
$\delta$ is the set of labels of $D$-vertices,

$\Lambda=\{(\pi(In(y)),\pi(y))\mid y\in Y^{\prime}\},$

$\Xi$ is the set of triples

$(\pi(x),\pi(Out(x)),\{(\pi(In(y)),\pi(y))\mid y\in Y^{\prime},\nexists
u.(x\rightarrow u\rightarrow y)\})$

\qquad\qquad for all $x\in X^{\prime}.$
\end{quote}

\bigskip

The accepting states are those such that $\Xi$\ contains a triple of the form
$(a,\Lambda,\emptyset)$. We show some transitions.

\begin{quote}
$\oplus\lbrack(\delta_{1},\Lambda_{1},\Xi_{1}),(\delta_{2},\Lambda_{2},\Xi
_{2})]\rightarrow(\delta_{1}\cup\delta_{2},\Lambda_{1}\cup\Lambda_{2},\Xi
_{1}\cup\Xi_{2}\cup\Xi^{\prime})$

where $\Xi^{\prime}$ is the set of triples are of the form $(a,\eta,\Phi
\cup\Lambda_{2})$ for $(a,\eta,\Phi)\in\Xi_{1}$ and of the form $(a,\eta
,\Phi\cup\Lambda_{1})$ for $(a,\eta,\Phi)\in\Xi_{2}$.
\end{quote}

\bigskip

The other transitions are easy to define, by the same methods as for the
previous automata. If $C\cup D$ is finite, $k=\left\vert C\right\vert $ and
$\ell=\left\vert D\right\vert $, the number of states is bounded by $2^{\ell
}.2^{k.2^{\ell}}$.$2^{k.2^{\ell}.2^{k.2^{\ell}}}$.\ The size of a state is
$O(2^{k.2^{\ell}})$ and the computation time of a transition is
$O(2^{k.2^{\ell}})$.

\bigskip

Table 4 compares the bounds on the sizes of the states for the linear FPT-FA
we just constructed for incidence graphs to the bounds for those constructed
in \cite{BCID12}\ for "ordinary" graphs. For terms obtained by Proposition 6
from a tree-decomposition of width $p$, we have $k=2$ and \ $\ell\leq2p+3.$

\begin{center}%
\begin{tabular}
[c]{|l|c|c|}\hline
Property & MSO & MSO$_{2}$\\\hline\hline
$edg$ & \multicolumn{1}{|l|}{$O(\log(k))$} & \multicolumn{1}{|l|}{$O(\log
(k)+\ell)$}\\\hline
$Link^{\exists\exists}$ & \multicolumn{1}{|l|}{$O(k)$} &
\multicolumn{1}{|l|}{$O(k+\ell)$}\\\hline
$Link^{\forall\exists}$ & \multicolumn{1}{|l|}{$O(k)$} &
\multicolumn{1}{|l|}{$O(k.2^{\ell})$}\\\hline
$Link^{\forall\forall}$ & \multicolumn{1}{|l|}{$O(k^{2})$} &
\multicolumn{1}{|l|}{$O(k.2^{\ell})$}\\\hline
$Link^{\exists\forall}$ & \multicolumn{1}{|l|}{$O(2^{k})$} &
\multicolumn{1}{|l|}{$O(2^{k.2^{\ell}})$}\\\hline
\end{tabular}

\bigskip

Table \ 4: Comparison between MSO and MSO$_{2}$-automata.
\end{center}

\section{Automata for other properties}

\bigskip

\subsection{$Inc$\emph{-}invariant properties}

\bigskip

\textbf{Definition 8} : \emph{Invariance for taking the incidence graph.}

A graph property $P$ is $Inc$-\emph{invariant} if $P(G)\Longleftrightarrow
P(Inc(G))$ for every graph $G$.

So are, for instance, for a directed graph $G,$ connectedness and strong
connectedness, the properties that all its vertices are of outdegree at most
$p$, that $G$ has a directed cycle, or an undirected cycle (a cycle in which
edge directions do not matter), or that $G$ has a path from vertex $x$ to
vertex $y$.

\bigskip

For such a property $P$, an FA over $F_{C,D}$ intended to check it on
incidence graphs defined by correct terms can be constructed from the FA
$\mathcal{A}_{P,C}$ over $F_{C}$ that checks $P$ on "ordinary" graphs $G,$
without needing significant modification: essentially, we replace $C$ by
$C\cup D.$ Although the FA for the relation $edg(X,Y)$ is more complicated in
the case of incidence graphs than for ordinary graphs, this increasing
complexity does not extend to all properties expressed by means of $edg(X,Y)$.

\bigskip

\subsection{An automaton for \textsc{Directed Hamiltonian cycle}}

\bigskip

The property $DirHam$ that a graph $G$ has a directed Hamiltonian cycle is
MSO$_{2}$ expressible but not MSO expressible. It is expressed in $Inc(G)$ by
"there exists a set of edges of $G$ that forms a directed cycle going through
all vertices". Without using the corresponding logical expression, we will
construct an FA\ for this property.\ 

\bigskip

We let $P$ mean be that the considered graph is a directed cycle or is empty,
and $L$ be the set of correct irredundant terms $t$ in $T(F_{C,D})$ such that
$P(val(t))$ holds.\ We first construct an FA\ $\mathcal{B}$ over $F_{C,D}$
that recognizes $L$ among\ correct and irredundant terms, that is, such that
$L(\mathcal{B})\cap L_{Irr}\cap L_{CT}=L.$

Let $t\in L$ and $t^{\prime}$ be a subterm of $t$.\ Then $val(t^{\prime})$
satisfies one of the following properties:

\begin{quote}
(a) it is a single directed cycle (and then $val(t^{\prime})=val(t)$),

(b) it consists of isolated vertices $x_{1},...,x_{p},(p\geq0)$ and of
pairwise disjoint paths from $y_{1}$ to $z_{1}$, $y_{2}$ to $z_{2}$, ...\ ,
$y_{m}$ to $z_{m}$ $(m\geq0)$, such that the labels of $x_{1},...,x_{p},$
$y_{1}$, $...,y_{m}$, $z_{1}$,...\ , $z_{m}$ are all different.
\end{quote}

Each of them implies :

\begin{quote}
(c) no $C$-vertex has indegree or outdegree 2 or more.
\end{quote}

\bigskip

Since $t$ is assumed correct, every $D$-vertex has indegree and outdegree at
most one.\ A transition that shows a violation of (c) will yield $Error$.

\bigskip

We define $\mathcal{B}$ with the following states : $Ok$, $Error$ and the
3-tuples $(\alpha,\beta,\Psi)$ in $\mathcal{P}_{f}(C\cup D)\times
\mathcal{P}_{f}(C)\times\mathcal{P}_{f}((C\cup D)^{2}).$ At the root of a term
$t^{\prime}\in T(F_{C,D})$ as above, the automaton $\mathcal{B}$ reaches the
following state:

\begin{quote}
$Ok$ if (a) holds,

$Error$ if neither (a) nor (b) holds,

$(\alpha,\beta,\Psi)$ if (b) holds and :

$\qquad\alpha$ is the set of labels of the vertices $x_{1},...,x_{p}$,

$\qquad\beta$ is the set of labels of the $C$-vertices of indegree and
outdegree 1,

$\qquad\Psi=\{(\pi(y_{1}),\pi(z_{1})),...,(\pi(y_{m}),\pi(z_{m}))\}.$
\end{quote}

From any finite set $\Psi=\{(a_{1},b_{1}),...,(a_{m},b_{m})\}\subseteq(C\cup
D)^{2}$, we define $\Psi_{1}:=\{a_{1},...,a_{m}\}$ and $\Psi_{2}%
:=\{b_{1},...,b_{m}\}$.

If $\sigma=(\alpha,\beta,\Psi)$ is reached at the root of $t^{\prime}$
satisfying (b), then:

\begin{quote}
(i) the sets $\alpha,\beta,\Psi_{1}$ and $\Psi_{2}$ are pairwise disjoint,

(ii) no two pairs in $\Psi$ have a component in common (by (i) this condition
reduces to: $a_{i}\neq a_{j}$ and $b_{i}\neq b_{j}$ for $i\neq j). $
\end{quote}

We denote by $D(\sigma)$ the conjunction of these two conditions.

\bigskip

\emph{Remarks}: When $\mathcal{B}$ reaches state $(\alpha,\beta,\Psi)$, the
labels of the $C$-vertices of $val(t^{\prime})$ are all in $\alpha\cup
\beta\cup\Psi_{1}\cup\Psi_{2}$, but $(\alpha,\beta,\Psi)$ does not indicate
the set $\delta$ of labels of the $D$-vertices that are on the paths from
$y_{j}$ to $z_{j}$ but not at their ends.\ These vertices have indegree and
outdegree 1. Since $t$ is assumed correct and irredundant, Lemma 2 yields that
$\delta\cap(\alpha\cup\Psi_{1}\cup\Psi_{2})=\emptyset.\square$

\bigskip

The accepting state is $Ok$. We now describe some transitions.

\begin{quote}
$\varnothing\rightarrow\overrightarrow{\emptyset},$

$\mathbf{a}\rightarrow(\{a\},\overrightarrow{\emptyset})$ for $a\in C\cup D,$

$\oplus\lbrack Ok,\overrightarrow{\emptyset}]\rightarrow Ok,$

$\oplus\lbrack Ok,q]\rightarrow Error,$ if $q\neq\overrightarrow{\emptyset}$
(in particular if $q=Ok$),

$\oplus\lbrack\sigma,\sigma^{\prime}]\rightarrow\sigma\cup\sigma^{\prime}$ if
$\alpha\cap\alpha^{\prime}=\emptyset,\Psi\cap\Psi^{\prime}=\emptyset$ and
$D(\sigma\cup\sigma^{\prime})$ holds,

where, if $\sigma=(\alpha,\beta,\Psi)$ and $\sigma^{\prime}=(\alpha^{\prime
},\beta^{\prime},\Psi^{\prime}),$ we have

$\sigma\cup\sigma^{\prime}:=$($\alpha\cup\alpha^{\prime},\beta\cup
\beta^{\prime},\Psi\cup\Psi^{\prime})$,

$\oplus\lbrack\sigma,\sigma^{\prime}]\rightarrow Error$ otherwise.
\end{quote}

\bigskip

We denote by $a,b$ any labels in $C\cup D$.\ The transitions for relabellings
$relab_{a\rightarrow b}$ \ in $F_{C,D}$ are:

\begin{quote}
$relab_{a\rightarrow b}[Ok]\rightarrow Ok,$

$relab_{a\rightarrow b}[\sigma]\rightarrow Error$ if $\{a,b\}\subseteq\alpha$
or $D(\sigma^{\prime})$ does not hold\footnote{That $D(\sigma^{\prime})$ holds
implies that $a$ and $b$ are not both in $\Psi_{1}$\ and not both in $\Psi
_{2}$.} \ 

where $\sigma^{\prime}$ is obtained from $\sigma$\ by replacing everywhere $a$
by $b$;

otherwise $\ relab_{a\rightarrow b}[\sigma]\rightarrow\sigma^{\prime}.$
\end{quote}

\bigskip

Let comment on the last case.\ If \ $a\in C$ and $a\notin\alpha\cup\beta
\cup\Psi_{1}\cup\Psi_{2}$, then $relab_{a\rightarrow b}$ has no effect and the
last transition yields $\sigma^{\prime}=\sigma.$ If \ $a\in D$ and
$a\notin\alpha\cup\Psi_{1}\cup\Psi_{2},$ it may happen that $a\in\delta$ (cf.
the above remarks) hence that $relab_{a\rightarrow b}$ might have some
effect.\ But since the input term is irrendundant and correct, Lemma 2 shows
that $b$ cannot belong to $\alpha\cup\Psi_{1}\cup\Psi_{2}$.\ Hence $\delta$ is
replaced by $\delta\lbrack a\rightarrow b]$ and nothing else is
modified.\ Hence the transition yields correctly $\sigma^{\prime}=\sigma.$

\bigskip

Transitions for edge additions are as follows:

$\overrightarrow{add}_{a\rightarrow b}[(\alpha,\beta,\Theta)]\rightarrow
\sigma$ if $a$ or $b$ is not in $\alpha\cup\beta\cup\Psi_{1}\cup\Psi_{2},$

$\overrightarrow{add}_{a\rightarrow b}[(\alpha,\beta,\Theta)]\rightarrow Ok$
if $\alpha=\emptyset$ and $\Psi=\{(b,a)\},$

$\overrightarrow{add}_{a\rightarrow b}[(\alpha,\beta,\Theta)]\rightarrow
(\alpha^{\prime},\beta^{\prime},\Psi^{\prime})$ as described in Table 5.

\begin{center}%
\begin{tabular}
[c]{|c|c|c|c|}\hline
Condition & $\alpha^{\prime}$ & $\beta^{\prime}$ & $\Psi^{\prime}%
$\\\hline\hline
\multicolumn{1}{|l|}{$a,b\in\alpha$} & \multicolumn{1}{|l|}{$\alpha-\{a,b\}$}
& \multicolumn{1}{|l|}{$\beta$} & \multicolumn{1}{|l|}{$\Psi\cup\{(a,b)\}$%
}\\\hline
\multicolumn{1}{|l|}{$a\in\alpha,(b,c)\in\Psi$} & \multicolumn{1}{|l|}{$\alpha
-\{a\}$} & \multicolumn{1}{|l|}{$\beta\cup\{b\}$} & \multicolumn{1}{|l|}{$\Psi
\cup\{(a,c)\}-\{(b,c)\}$}\\\hline
\multicolumn{1}{|l|}{$(d,a)\in\Psi,b\in\alpha$} & \multicolumn{1}{|l|}{$\alpha
-\{b\}$} & \multicolumn{1}{|l|}{$\beta\cup\{a\}$} & \multicolumn{1}{|l|}{$\Psi
\cup\{(d,b)\}-\{(d,a)\}$}\\\hline
\multicolumn{1}{|l|}{$(d,a),(b,c)\in\Psi$} & \multicolumn{1}{|l|}{$\alpha$} &
\multicolumn{1}{|l|}{$\beta\cup\{a,b\}$} & \multicolumn{1}{|l|}{$\Psi
\cup\{(d,c)\}-\{(d,a),(b,c)\}$}\\\hline
\end{tabular}

Table 5\ : Transitions $\overrightarrow{add}_{a\rightarrow b}[(\alpha
,\beta,\Psi)]\rightarrow(\alpha^{\prime},\beta^{\prime},\Psi^{\prime}).$
\end{center}

By Properties (i) and (ii), $a,b,c,d$ are pairwise distinct. Finally,

\begin{quote}
$\overrightarrow{add}_{a\rightarrow b}[(\alpha,\beta,\Psi)]\rightarrow Error$
in all other cases.
\end{quote}

Note that by our usual argument using Lemma 2, if $\overrightarrow
{add}_{a\rightarrow b}[(\alpha,\beta,\Psi)]$\ is to be fired, then $b$ cannot
belong to $\delta$.

If $C\cup D$ is finite, $k=\left\vert C\right\vert $ and $\ell=\left\vert
D\right\vert $, then the number of states is bounded by $3^{k}.2^{\ell
}.m(k+\ell)$ where $m(n)$ is the number of ordered matchings over $n$ elements
(of sets of pairwise disjoint ordered pairs). We have $\ \left\lfloor
n/2\right\rfloor !<m(n)<2^{n^{2}}$.\ The size of a state is $O((k+\ell
)\log(k+\ell))$ \ and the time for computing a transition is also
$O((k+\ell)\log(k+\ell))$.

The property $DirHam$, i.e., the existence of a directed Hamiltonian cycle in
the graph $G$ such that $Inc(G)=val(t)$ is expressed by :

\begin{quote}
"There exists a set $U$ of $D$-vertices of $val(t)$

such that $P(val(t)[V_{val(t)},U])$ holds"
\end{quote}

(cf.\ the end of Section 3).\ Note that $t[V_{val(t)},U]$ is a correct
irredundant term for all sets $U$ if $t$ is so. This property is thus checked
by the nondeterministic FA $\mathcal{C}$ obtained from $\mathcal{B}$ by adding
the transitions $\mathbf{d}\rightarrow\overrightarrow{\emptyset}$ for $d\in
D.$ These new transitions correspond to the elimination of the edges that will
not be on the directed cycle under construction. The FPT-FA $\det
(\mathcal{C)}$ checks $DirHam$ in time $O(2^{2(k+\ell)^{2}}.n)$ where $n$ is
the number of vertices and edges of the input graph.

\bigskip

\section{Conclusion}

\bigskip

These results indicate that the tools of \cite{BCID12, BCID13} can be applied
to the verification of MSO$_{2}$ properties of graphs of bounded tree-width
given by their tree-decompositions. The software AUTOGRAPH\footnote{The system
AUTOGRAPH that builds and runs FA is written in LISP \cite{BCID11} and
http://dept-info.labri.u-bordeaux.fr/\symbol{126}idurand/autograph} can be
used basically as it is (up to minor syntactic adaptations) although the
algebras of terms describing tree-decompositions and of terms defining
clique-width are fairly different (as discussed in \cite{Cou12}).

\bigskip

We have introduced in \cite{Cou12} a variant of tree-width called
\emph{special tree-width} intended for the verification of MSO$_{2}$
properties.\ It is weaker than tree-width in the sense that bounded special
tree-width implies bounded tree-width but not vice-versa.\ Its advantage is
that \emph{special tree-decompositions} can be formalized in terms of
clique-width operations.\ A decomposition of width $p$ is formalized by a term
in $T(F_{[p+2]})$.\ Hence, this article also provides tools for checking
MSO$_{2}$ properties by FA based on clique-width operations.

For completeness, we also cite \cite{KLR} where a completely different method
is used to check MSO$_{2}$ properties of graphs of bounded tree-width.

\bigskip

\emph{Acknowledgements }: I thank R.\ Sampaio for fruitful discussions on
these topics.\ I thank I.\ Durand for her ongoing collaboration and her
implementation of fly-automata.

\newpage

\end{document}